\documentclass[twocolumn,amsmath,amssymb,pra,aps,superscriptaddress]{revtex4}

\usepackage[pdftex]{graphicx}
\usepackage{bm}
\usepackage{comment}
\usepackage{mathrsfs}
\usepackage{here}
\usepackage{amsmath}
\usepackage{amssymb}
\usepackage[utf8]{inputenc}
\usepackage{color}
\usepackage{ascmac}
\usepackage{amsfonts}

\usepackage{ulem}
\usepackage{eucal}
\usepackage{framed}
\usepackage{cancel}
\usepackage{braket}

\definecolor{shadecolor}{cmyk}{0.08,0.05,0.0,0.0}

\newcommand{\I}{{\mathrm{i}}}
\newcommand{\E}{{\mathrm{e}}}
\newcommand{\D}{{\mathrm{d}}}

\newcommand{\Blue}[1]{{#1}}

\newcommand{\bq}{\begin{eqnarray}}
\newcommand{\eq}{\end{eqnarray}}

\newcommand{\au}[1]{\if0{#1}\fi}  % hide atomic unit

\newcommand{\siu}[1]{#1} % show SI

\begin{document}

\title{Low-energy scattering of ultracold atoms by a dielectric nanosphere}

\author{T.~Yamaguchi}
\affiliation{Department of Accelerator Science, The Graduate University for Advanced Studies (SOKENDAI), 1-1 Oho, Tsukuba, Ibaraki 305-0801, Japan}
\author{D.~Akamatsu}
\affiliation{Department of Physics, Gratuate School of Engineering Science, Yokohama National University, 79-5 Tokiwadai, Hodogaya-ku, Yokohama, Kanagawa 240-8501, Japan}
\affiliation{Precursory Research for Embryonic Science and Technology (PRESTO), Japan Science and Technology Agency (JST), Kawaguchi, Saitama
332-0012, Japan}
\author{R.~Kanamoto}
\affiliation{Department of Physics, Meiji University, Kawasaki, Kanagawa 214-8571, Japan}
% ___________________________________________________________________________________________
\begin{abstract}
We theoretically study the low-energy scattering of ultracold atoms by a dielectric nanosphere of silica glass levitated in a vacuum. The atom and dielectric surface interact via dispersion force of which strength sensitively depends on the polarizability, dielectric function, and geometry. For cesium and rubidium atoms, respectively, we compute the atom-surface interaction strength, and characterize the stationary scattering states by taking adsorption of the atoms onto the surface into account. As the energy of the incoming atoms is lowered, we find that differences between quantum and classical scatterings emerge in two steps. 
First, the quantum-mechanical differential cross section of the elastic scattering starts to deviate from the classical one 
at an energy scale comparable to a few microkelvin in units of temperature due to the de Broglie matter-wave diffraction. 
Second, the differences are found in the cross sections in a regime lower than a few nanokelvin, where the classically forbidden reflection occurs associated with the $s$-wave scattering, and the discrete nature of angular momentum. 
We also study the dependencies of quantum and classical scattering properties on the radius of the nanosphere. This paper paves the way to identify the quantum regime and to understand the physical origin of quantum effects in the collisions between a nanoparticle and environmental gas over various temperatures. 
\end{abstract}

\maketitle

% ___________________________________________________________________________________________
\section{Introduction}

There has been growing interest in optically levitated nanoparticles~\cite{Neukirch} because of their potential applicabilities to unexplored fields  
such as non-equilibrium dynamics and thermodynamics at the nanoscale~\cite{Millen2014}, suspension-free ultra high-Q optomechanics~\cite{Chang2010, Kiesel2013, Kamba}, studies of quantum-classical boundaries~\cite{Romero-Isart2011, Bateman2014, Toros}, nonlinear dynamics~\cite{Pettit, Gieseler2014}, 
weak-force sensing~\cite{Jain, Geraci2010, Arvanitaki}, and control of translational, librational~\cite{Liu2017}, rotational~\cite{Reimann2018, MB, Nishikawa}, and precessional~\cite{Rashid2018} motions. 
Experimentalists have achieved ultimate quantum control and ground-state cooling of the center-of-mass motion of a nanoparticle with measurement and feedback~\cite{Magrini, Tebbenjohanns} that were thus far successful only in atomic and optical sciences. 
It also provides an opportunity to explore cavity QED effects in thermal radiation of isolated nano- or micron-sized objects~\cite{Odashima, Morino}.

A levitated nanoparticle interacts with the background gas in addition to the optical field that enables trapping and manipulation. The interaction between a neutral atom or molecule and polarizable materials is called dispersion force, which arises from instantaneous fluctuations of the dipoles in the relevant matters~\cite{Buhmann}. 
Nonetheless, the collisional properties between the nanoparticle and the background gas depend little on the details of 
the force at room temperature, and the nanoparticle undergoes intense Brownian motions due to the random momentum kicks imparted by collisions. 
The collisions are, in contrast, expected to be qualitatively altered at low temperatures where the de Broglie wave character of surrounding atoms or molecules manifests itself. In such low temperatures, atom-atom and atom-molecule scatterings have been widely studied in the context of ultracold chemistry~\cite{Idziaszek, Micheli, Julienne, Jachymski}. With similar theoretical approaches established in the studies of ultracold chemistry,  the scattering of ultracold sodium and metastable helium (2 $^3$S) atoms by an absorbing nanosphere of conducting material has been studied in Ref.~\cite{Arnecke}, where a form of dispersion-force potential between the atom and spherical surface is determined, and various scattering cross sections are obtained as functions of the incoming energy.  However, the regime investigated in Ref.~\cite{Arnecke} was limited within the extremely 
low energies of the order of nanokelvin in units of temperature where the quantum reflection~\cite{Carraro} of atoms is predicted as a purely quantum effect. This energy is far below an energy scale where the de Broglie wavelength of the atom is comparable to the typical size of the nanosphere. Therefore, it is still elusive at what energy scale the atom-nanoparticle scattering starts to be sensitive to the details of the force, and for which observables the classical picture of collisions is invalid.

The aim of this paper is to identify quantum effects in the scattering of cesium and rubidium atoms by a dielectric nanosphere.  For this purpose, we explore the wide range of the atomic incoming energy from the high-energy limit to the $s$-wave scattering regime by comparing several classical and quantum scattering observables. The radius of the nanosphere is also varied for a clarification of finite-size effects that are absent in atom-molecule scattering. We determine the atom-surface potential strengths by employing the formulation used in Ref.~\cite{Arnecke}. For inclusion of adsorption of atoms onto the surface, we utilize the semiclassical boundary condition for the atomic wavefunction, which has been also established to describe the reactive process in ultracold molecules~\cite{Idziaszek, Micheli, Julienne, Jachymski} and quantum reflection~\cite{Cote96, Cote97, Cote98}. 
As the energy of the incident atomic beam is lowered, we found that the first quantum effect reveals in the differential cross section of the elastic scattering at an energy scale of a few microkelvin in units of temperature, where the thermal de Broglie wavelength of the atom is comparable to the size of the nanosphere. 
In contrast, the quantum and classical absorption cross sections and loss rates agree quite well, even at much lower energies. The second set of quantum effects associated with the $s$-wave scattering, such as quantum reflection, emerge in a regime lower than a few nanokelvin, as manifestations of the discreteness of the quantum-mechanical angular momentum. 
We also found numerically that the scattering length depends on the radius of the nanosphere in a nontrivial manner.

The knowledge of the interaction between atoms and a dielectric material provides useful information to state-of-the-art experiments of levitated nanoparticles~\cite{Chang2010, Kiesel2013, Bateman2014, Pettit, Geraci2010, Arvanitaki, Liu2017, Reimann2018, Magrini, Tebbenjohanns}. The interface of ultracold atoms and solid-state materials interacting via dispersion force 
has also drawn attention~\cite{Schneeweiss, Jetter} due to the crucial importance for fabrication and control of nanoscale devices~\cite{Hummer}. Furthermore, it may provide the possibility of substituting the roles of photons in optomechanics  with coherent matter waves such that atoms are utilized to control and sense the Brownian motion of mechanical oscillators 
by exploiting the collective enhancement due to massive coherent atoms in the sprit of the matter-wave optics~\cite{Deppner}.

This paper is organized as follows. 
In Sec.~\ref{interaction}, we revisit the construction of the dispersion-force potential between a neutral atom and spherical surface, and numerically determine the potential strengths for the case of cesium, rubidium atoms, and a dielectric nanosphere of silica glass. The classical scattering by this potential field is studied in Sec~\ref{sec:classical_scat}. In Sec.~\ref{st}, we formulate the quantum-mechanical scattering, and introduce an absorption ansatz of the incident matter wave by imposing a boundary condition in the proximity of the surface. The numerical results of low-energy quantum-mechanical scattering, in particular, the incident atomic energy and nanospherical size dependencies, are shown in Sec.~\ref{results}. Section~\ref{sum} summarizes and concludes our results.

% ___________________________________________________________________________________________
\section{Dispersion forces}\label{interaction}

\begin{figure}[t]
\centering
%\aps{\includegraphics[width=8cm,pagebox=cropbox]{fig_1.pdf}}
\includegraphics[width=8cm]{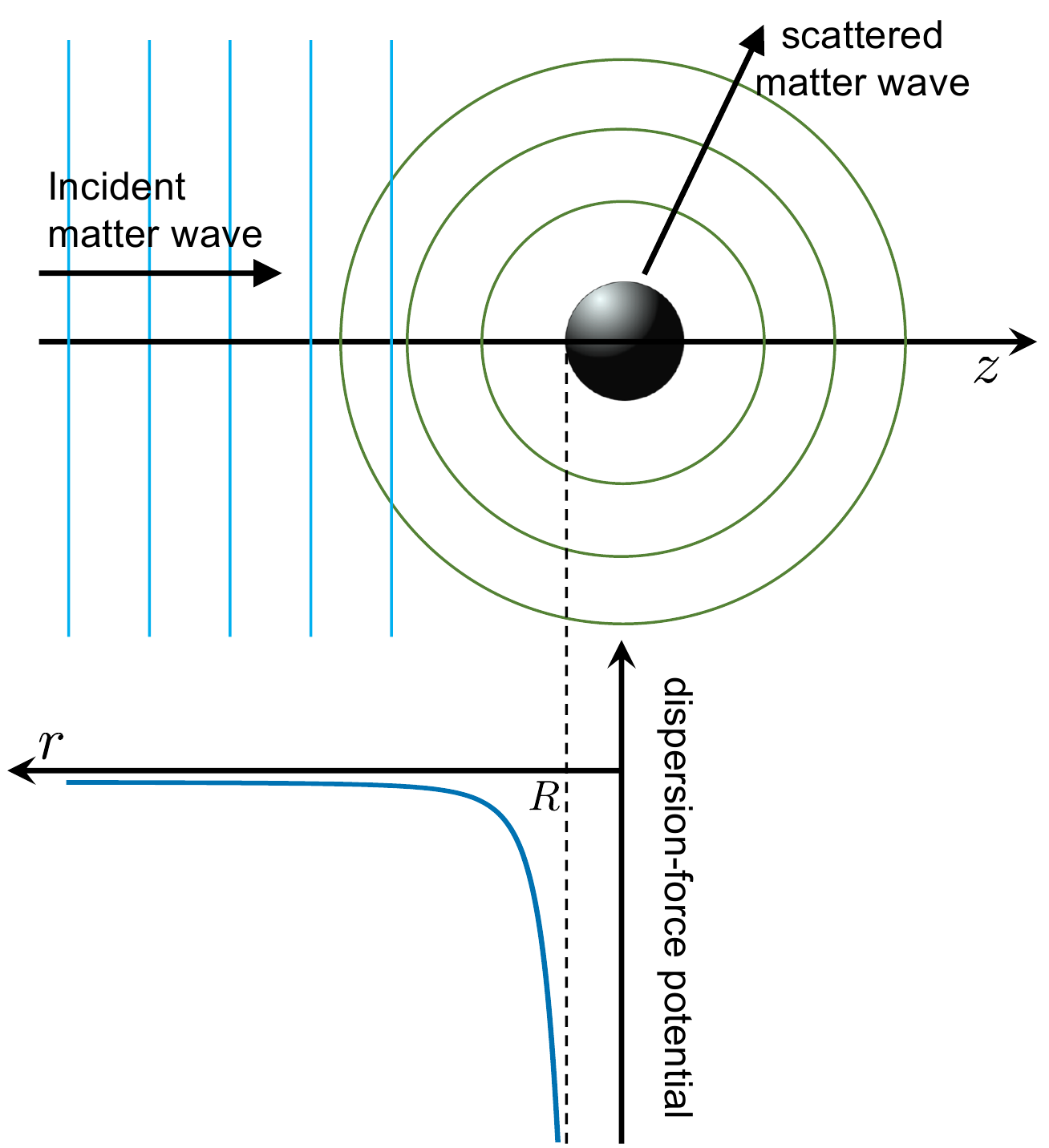}
\caption{Plane matter wave of energy $E$ propagating along the $z$-axis, is incident on a dielectric nanosphere of radius $R$, and is scattered by the atom-surface interaction potential. 
This potential gives rise to the dispersion force, which is central and attractive. Close to the surface, the force is deeply attractive.}
\label{fig:system}
\end{figure}

We consider a situation as shown in Fig.~\ref{fig:system}, where the plane wave of an ultracold atom with energy $E$ 
(or $2E/k_B$ in units of temperature) is incident on a dielectric nanosphere of radius $R$, and is scattered by the atom-surface potential. 
Hereafter, every order of magnitude of temperature refers to this converted incoming energy into temperature, which provides a good measure of energy scale in various experiments in atomic, molecular, and optical sciences, though it does not correspond to 
the thermodynamic temperature, nor involve any averaging over an energy distribution. 
Throughout this paper, we also assume that the nanosphere experiences no recoil associated with the collisions of atoms. 
In this section, we determine the interaction potential between a cesium or rubidium atom and the spherical surface of silica glass. 

% ___________________________________________________________________________________________
\subsection{Construction of atom-surface potential}
%--------------------------------
%: 【figure】 system
%--------------------------------

The precise form of atom-surface potential is sensitive to various parameters such as characteristic wavelengths of atomic transitions, as well as the shape, dielectric function, and polabizability of the material. 
The exact treatments of dispersion interaction between an atom and a spherical surface for an arbitrary distance in terms of macroscopic QED are found in Ref.~\cite{Buhmann}. In this paper, we nonetheless employ an approximate construction of the dispersion-force potential~\cite{Arnecke} for numerical simplicity and analytical transparency. The exact and approximate potentials will be quantitatively compared later in this section. 

%: atom-flat-surface
Let $R$, $r$, and $r' = r-R$ be the radius of a nanosphere, the distance of the atom from the center of the sphere, and distance from the surface of sphere, respectively. In this paper, we denote all the relevant formula in the International System of Units 
(SI units) unless otherwise stated, while we perform all the numerical calculations in atomic units (a.u.). 

When the atom-surface distance $r'$ is much smaller than the radius $R$ of the sphere, 
the interaction potential is well approximated with the one $V_{\rm fw}(r')$ between the atom and a semi-infinite {\it flat wall}. The expression for $V_{\rm fw}(r')$ was introduced by Tikochinsky and Spruch~\cite{Tikochinsky} as
\au{
\begin{align}
	V_{\rm fw}(r') &= - \frac{\alpha_{\rm fw}^3}{2 \pi} \int_0^\infty \D \xi \xi^3 \alpha(\I \xi) \nonumber\\
&\qquad\qquad \times \int_1^\infty \D p\, \E^{-2 \alpha_{\rm fw} \xi p r'} H(p, \epsilon(\I \xi)), \label{flat}
\end{align}
}
\siu{
\begin{align}
	V_{\rm fw}(r') &= - \frac{\hbar}{8\pi^2\epsilon_0 c^3} \int_0^\infty d \xi \xi^3 \alpha(\I \xi) \nonumber\\
&\qquad\qquad \times \int_1^\infty d p\, e^{-2 \xi p r'/c} H(p, \epsilon(\I \xi)), \label{flat}
\end{align} 
}
where \au{$\alpha_{\rm fw}$ is the fine structure constant and}
\begin{align}
&H(p, \epsilon) = \frac{\sqrt{\epsilon-1 + p^2} - p}{\sqrt{\epsilon - 1 + p^2} + p} \nonumber\\
&\qquad\qquad\quad+ (1 - 2p^2) \frac{\sqrt{\epsilon - 1 + p^2} - \epsilon p}{\sqrt{\epsilon - 1 + p^2} + \epsilon p}, 
\end{align}
and $\epsilon_0$ and $c$ are the permittivity of vacuum and the speed of light, respectively. 
The quantities $\alpha(\I \xi)$ and $\epsilon(\I \xi)$ denote the atomic polarizability and the dielectric function of the surface at imaginary frequencies, respectively.

%: atom-infinitesimal-sphere

When the atom is far from the spherical surface, $r' \gg R$, on the other hand, the potential is described by the one $V_\mathrm{pp}(r)$ between the atom and a {\it point particle}~\cite{Buhmann}:
\au{
\begin{align}
V_\mathrm{pp}(r) = -\frac{1}{\pi r^6} & \int_0^\infty \D \xi \alpha(\I \xi) \alpha_\mathrm{sphere}(\I \xi) g(\xi r/c), \label{pot_pp}
\end{align}
}
\siu{
\begin{align}
V_\mathrm{pp}(r) = -\frac{\hbar}{16\pi^3 \epsilon_0^2 r^6} & \int_0^\infty d \xi \alpha(\I \xi) \alpha_\mathrm{sphere}(\I \xi) g(\xi r/c), \label{pot_pp}
\end{align}
}
where $g(x) = e^{-2 x}  (3 + 6x + 5x^2 + 2x^3 + x^4)$, and 
\au{
\begin{align}
\alpha_\mathrm{sphere} (\I \xi) = R^3 \frac{\epsilon(\I \xi) - 1}{\epsilon(\I \xi) + 2}
\end{align}
}
\siu{
\begin{align}
\alpha_\mathrm{sphere} (\I \xi) = 4\pi \epsilon_0 R^3 \frac{\epsilon(\I \xi) - 1}{\epsilon(\I \xi) + 2}
\end{align}
}
is the total polarizability of the dielectric sphere~\cite{Jackson}. 

%: non-retarded and retarded regimes

In addition to the radius $R$ of the nanosphere, the wavelength $\lambda$ of the dominant atomic transition is also an important length scale that characterizes the interaction potential.  
For rubidium and cesium atoms, the typical wavelength $\lambda$ is within a range of 700--900 nm. 
In a dispersion-force potential, the retardation effect of the electromagnetic field is significant when $r' \gg \lambda$. 
This regime is thus called the retarded regime, as discussed by Casimir and Polder~\cite{Casimir}. The retardation effects in the atom-atom potential of the form Eq.~(\ref{pot_pp}) was studied in Ref.~\cite{Marinescu}. In contrast, the retardation plays no role when $r' \ll \lambda$, which is called the non-retarded regime. Each potential behaves in the non-retarded and retarded limits as 
\begin{align}
V_{\rm fw}(r') &\xrightarrow{r' \ll \lambda} - \frac{C_3}{r'^3}, \quad V_{\rm fw}(r') \xrightarrow{r' \gg \lambda} - \frac{C_4}{r'^4}, \\
V_\mathrm{pp}(r) & \xrightarrow{r' \ll \lambda} - \frac{C_6}{r^6}, \quad V_\mathrm{pp}(r) \xrightarrow{r' \gg \lambda} - \frac{C_7}{r^7}. 
\end{align}

%: entire potential

Taking these geometric and internal length scales into account, we employ 
the following form of the interaction potential between an atom and the spherical surface~\cite{Arnecke}: 
\begin{align}
	V(r) = - \frac{\hbar^2}{2 \mu} \left.
\left[ \frac{r'^3}{\beta_3}v\left(\frac{r'}{L}\right) + \frac{r'^6}{\beta_6^4}v\left(\frac{r'}{L'}\right) \right]^{-1} \right|_{r' = r-R},\label{Vint}
\end{align}
where $\mu$ is the reduced mass that can always be replaced with the atomic mass,  
and 
\begin{align}
\beta_n = \left(\frac{2 \mu C_n}{\hbar^2} \right)^{1/(n-2)} \qquad (n = 3, 4, 6, 7) 
\end{align}
is a length parameter as a measure of the typical interaction range. The parameters 
\begin{align}
 \,  L = \frac{\beta_4^2}{\beta_3},
\quad L' = \frac{\beta_7^5}{\beta_6^4}
\end{align}
characterize the typical length scales at which the power law crosses over from non-retarded to retarded behaviors in $V_{\rm fw}$ and $V_{\rm pp}$, respectively. 
The function $v(x)$ is an arbitrary form that satisfies asymptotic behaviors
\begin{align}
	\lim_{x \to 0} \frac{1}{v(x)} = 1, \quad \lim_{x \to \infty} \frac{1}{v(x)} = \frac{1}{x}, 
\end{align}
and smoothly connects different power-law dependencies in the potential. 
In this paper, we use an expression~\cite{Shimizu2001}
\begin{align}\label{eq:shapef}
	v(x) = 1 + x. 
\end{align}
In the following analysis, we employ the form Eq.~\eqref{Vint} with the shape function Eq.~\eqref{eq:shapef} as the interaction potential between an atom and spherical surface. By definition, the potential in the proximity of the surface and at a large distance from the surface, respectively, behaves as 
\bq
V(r) \xrightarrow{r \to R} -\frac{C_3}{(r-R)^3},\quad 
V(r) \xrightarrow{r \to \infty}  -\frac{C_7}{r^7}.
\eq

% ___________________________________________________________________________________________
\subsection{Coefficients $C_n$}

%--------------------------------
%: 【table】 parameters
\begin{table*}
	\centering
	\caption{Known values of static polarizabilities $\alpha (0)$, interatomic van der Waals coefficients $C'_\mathrm{a}$ for rubidium and cesium, and obtained characteristic frequencies $\omega_\mathrm{a}$. The static dielectric constant of silica glass $\epsilon (0)$ and the corresponding value of the function $\phi(\epsilon (0))$ are also tabulated.}
	\begin{tabular}{lccc}\hline
		\multicolumn{1}{c}{Parameter} &\multicolumn{1}{c}{~~~~~Value (a.u.)} & \multicolumn{1}{c}{~~~~~Value (SI)}&\multicolumn{1}{c}{~~~~~Ref.} \\
		\hline\hline 			
		$\alpha_\mathrm{Rb}(0)$ &~~~~~ 319.1 $\pm$ 6.4 & $~~~~~~~~ (5.26 \pm 0.11) \times 10^{-39}$ C$\cdot$m$^2$$\cdot$V$^{-1}$ & ~~~~~\cite{Miller} \\ % modified
		$\alpha_\mathrm{Cs}(0)$ &~~~~~ 401.0 $\pm$ 0.6 & $~~~~~~~~(6.611 \pm 0.009) \times 10^{-39}$ C$\cdot$m$^2$$\cdot$V$^{-1}$ & ~~~~~\cite{2003Amini-Gould} \\  % modified
		$C'_\mathrm{Rb}$ &~~~~~ 4691&  $~~~~~~~~4.49 \times 10^{-76}$ J$\cdot$m$^6$&~~~~~ \cite{Derevianko} \\
		$C'_\mathrm{Cs}$ &~~~~~ 6851&  $~~~~~~~~6.56 \times 10^{-76}$ J$\cdot$m$^6$&~~~~~ \cite{Derevianko}  \\
		$\omega_\mathrm{Rb}$ &~~~~~$6.139\times10^{-2}$&~~~$2\pi \times 0.40\times10^{15}\ \mathrm{s}^{-1}$ & ~~~~~Present \\
		$\omega_\mathrm{Cs}$ &~~~~~$5.681\times10^{-2}$&~~~$2\pi \times 0.37\times10^{15}\ \mathrm{s}^{-1}$ & ~~~~~Present \\
		\hline
		 $\epsilon (0)$ & \multicolumn{2}{c}{3.81}  &~~~~~\cite{CRC} \\
		$\phi(\epsilon(0))$ & \multicolumn{2}{c}{0.769}  &~~~~~Present \\\hline
	\end{tabular}
	\label{valuestab}
\end{table*}
%--------------------------------

We next compute the coefficients $C_n\ (n = 3, 4, 6, 7)$ appearing in the atom-surface interaction potential $V (r)$.  
The coefficients $C_n$ are expressed in terms of the atomic polarizability $\alpha$ and 
dielectric function $\epsilon$ of the sphere as~\cite{Buhmann}
\au{
\begin{align}
\begin{aligned}
	C_3 &= \frac{1}{4\pi} \int_0^\infty \alpha(\I \xi) \frac{\epsilon(\I \xi)-1}{\epsilon(\I \xi) +1} \D \xi  \\
	C_4 &= \frac{3}{8 \pi \alpha_{\rm fw}} \alpha(0) \frac{\epsilon(0)-1}{\epsilon(0)+1} \phi(\epsilon(0)) \\
	C_6 &= \frac{3R^3 }{\pi} \int_0^\infty \alpha(\I \xi) \frac{\epsilon(\I \xi)-1}{\epsilon(\I \xi) + 2} \D \xi \\
	C_7 &= \frac{23 R^3}{4\pi \alpha_{\rm fw}} \alpha(0)  \frac{\epsilon(0)-1}{\epsilon(0)+2}. 
\end{aligned} \label{vdWcoefficients}
\end{align}
}
\siu{
\begin{align}
\begin{aligned}
	C_3 &= \frac{\hbar}{16\pi^2 \epsilon_0} \int_0^\infty \alpha(\I \xi) \frac{\epsilon(\I \xi)-1}{\epsilon(\I \xi) +1} d \xi  \\
	C_4 &= \frac{3 \hbar c}{32 \pi^2 \epsilon_0} \alpha(0) \frac{\epsilon(0)-1}{\epsilon(0)+1} \phi(\epsilon(0)) \\
	C_6 &= \frac{3\hbar R^3 }{4 \pi^2 \epsilon_0}\int_0^\infty \alpha(\I \xi) \frac{\epsilon(\I \xi)-1}{\epsilon(\I \xi) + 2} d \xi \\
	C_7 &= \frac{23 \hbar c R^3}{16\pi^2 \epsilon_0} \alpha(0)  \frac{\epsilon(0)-1}{\epsilon(0)+2}. 
\end{aligned} \label{vdWcoefficientsSI}
\end{align}
}
The function $\phi(\epsilon)$ appearing in $C_4$ is defined as \cite{Yan}
\begin{align}
	\phi(\epsilon) &= \frac{\epsilon+1}{2(\epsilon-1)} \int_0^\infty \frac{1}{p^4}H(p, \epsilon) \D p \nonumber \\
	&= \frac{\epsilon+1}{\epsilon-1} \left[ \frac{1}{3} + \epsilon + \frac{4 - (\epsilon+1)\epsilon^{1/2}}{2(\epsilon-1)} + a(\epsilon) + b(\epsilon) \right] , 
\end{align}
where 
\begin{align}
	a(\epsilon) &= - \frac{\sinh^{-1}\left[(\epsilon-1)^{1/2}\right]}{2 (\epsilon-1)^{3/2}} \left[ 1 + \epsilon + 2\epsilon (\epsilon-1)^2 \right], \nonumber\\
	b(\epsilon) &= \frac{\epsilon^2}{(\epsilon+1)^{1/2}} \left[ \sinh^{-1} (\epsilon^{1/2}) - \sinh^{-1} (\epsilon^{-1/2}) \right] .
\end{align}

The coefficients $C_4$ and $C_7$, which describe the retarded behaviors of $V_{\rm fw}$ and $V_\mathrm{pp}$, respectively, are determined by the static polarizability $\alpha (0)$ of the atom and the static dielectric constant $\epsilon(0)$ of the sphere. The known values of the static polarizabilities of rubidium~\cite{Miller} and cesium atoms~\cite{2003Amini-Gould}, and of the dielectric constant of silica glass~\cite{CRC}, are tabulated in Table~\ref{valuestab}. 

For the integrals in $C_3$ and $C_6$, which describe the non-retarded behaviors of $V_{\rm fw}$ and $V_\mathrm{pp}$, respectively, we need the atomic polarizability and the dielectric function of silica glass at imaginary frequencies. These are evaluated by the one-oscillator model and Lorentz model as follows.

%: atomic polarizability
\subsubsection{Atomic polarizability}

We employ the one-oscillator model \cite{Oberst, Vidali} for the expression of the atomic polarizability, 
\begin{align}
%%	\alpha(\I \xi) = \frac{\alpha(0)}{1 + \displaystyle \left(\frac{\xi}{\omega_\mathrm{a}} \right)^2}, 
\alpha(\I \xi) = \frac{\alpha(0)}{1+(\xi/\omega_{\rm a})^2}
\label{one-oscillator}
\end{align}
where $\omega_\mathrm{a}$ is the characteristic frequency of an atom, determined by the van der Waals interaction coefficient $C'_\mathrm{a}$ between the identical atoms~\cite{Buhmann}
\begin{align}
	C'_\mathrm{a} = \frac{3\hbar}{16\pi^3 \epsilon_0^2} \int_0^\infty \alpha(\I \xi)^2 d \xi .\label{CA}
\end{align}
Substituting Eq.~\eqref{one-oscillator} into Eq.~\eqref{CA}, the characteristic frequency is obtained as follows~\cite{Vidali}: 
\au{
\begin{align}
	\omega_\mathrm{A} = \frac{4 C'_\mathrm{A}}{3 \alpha^2 (0)}. \label{omegaA}
\end{align}
}
\siu{
\begin{align}
	\omega_\mathrm{a} = \frac{64 \pi^2 \epsilon_0^2 C'_{\rm a}}{3 \hbar \alpha^2(0)}. \label{omegaASI}
\end{align}
}
The values of $C'_\mathrm{a}$ for alkali-metal atoms have been evaluated in Ref.~\cite{Derevianko}. Using known values of the static polarizabilities $\alpha (0)$ and $C'_\mathrm{a}$, the characteristic frequency $\omega_\mathrm{a}$ is calculated by means of Eq. (\ref{omegaASI}) for rubidium and cesium atoms. The results are tabulated in Table \ref{valuestab}.

%: dielectric function of silica glass

%--------------------------------
%: 【figure】 dielectric function
\begin{figure}[b]
\centering
%\aps{\includegraphics[width=8.7cm,pagebox=cropbox]{fig_2.pdf}}
\includegraphics[width=8.7cm]{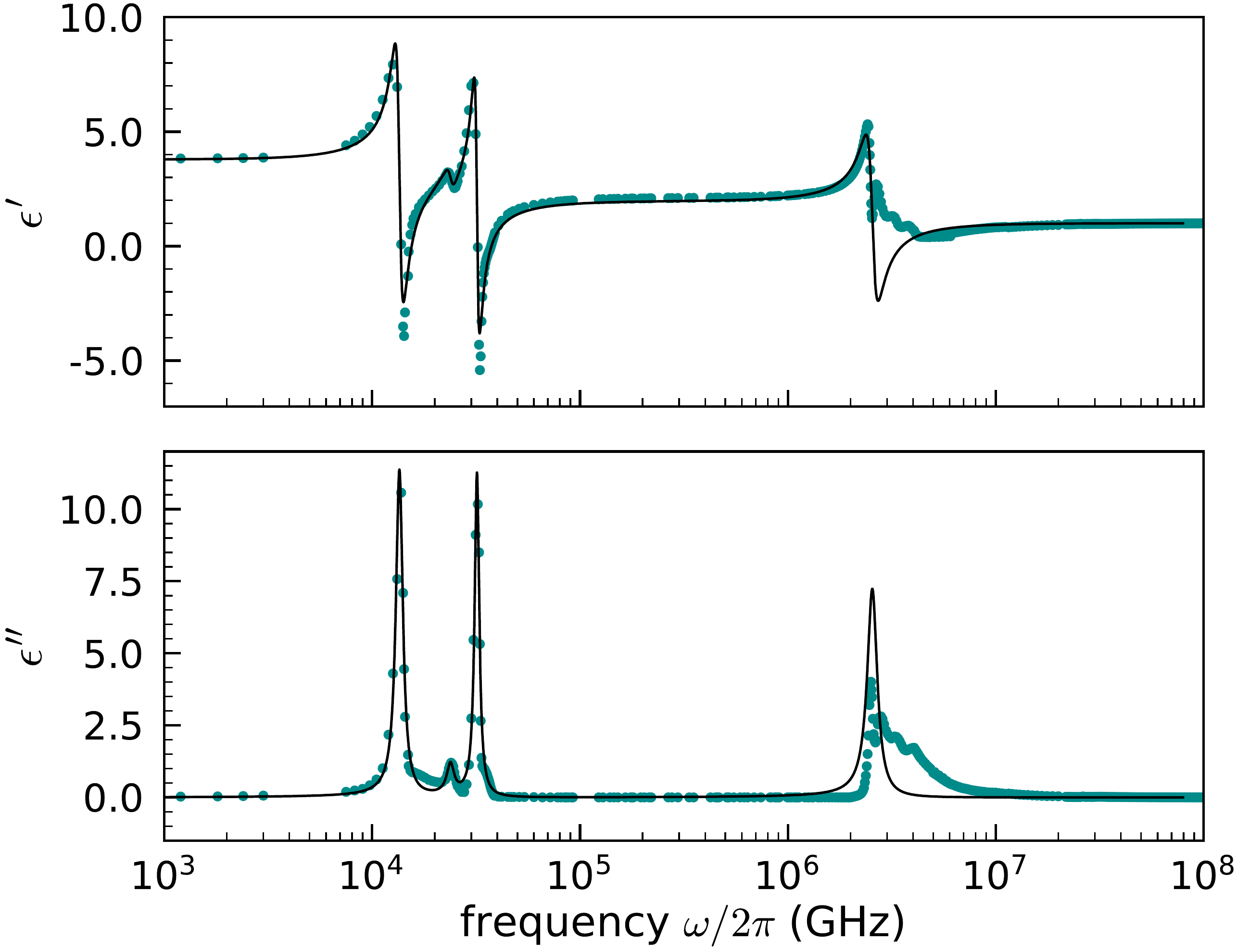}
\caption{Real (upper panel) and imaginary (lower panel) parts of the experimental data (dots) of the dielectric function of silica glass in Ref.~\cite{Palik} and that (solid curves) obtained by the fitting to the Lorentz model with optimal parameters.}
\label{fit fig}
\end{figure}
%--------------------------------

\subsubsection{Dielectric function of nanosphere}

We next address the dielectric function of silica glass at imaginary frequencies. In a wide range of frequencies, the dielectric function is approximated by the Lorentz model \cite{Fox}, 
\begin{align}
	\epsilon(\omega) %= \epsilon'(\omega) + \I \epsilon''(\omega) 
	= 1 + \sum_j \frac{ \tilde{\omega}_j^2}{\omega_j^2 - \omega^2 - \I \gamma_j \omega}, \label{Lorentz eq}
\end{align} 
where $\omega_j,\ \tilde{\omega}_j$, and $\gamma_j$ are the resonant frequencies, plasma frequencies, and damping rates of the oscillator, respectively. The real part $\epsilon' (\omega)$ and imaginary part $\epsilon'' (\omega)$ of the dielectric function are related to the refractive index and absorption coefficients, respectively. 

The experimental data of the complex reflectivity $\sqrt{\epsilon (\lambda)}$ of silica glass are tabulated in Ref. \cite{Palik}. 
We fit these data to Eq.~\eqref{Lorentz eq} in the frequency domain to obtain the optimal parameters $\tilde{\omega}_j,\ \omega_j$, and $\gamma_j$, and the results are tabulated in Table \ref{fit param}.

%--------------------------------
%: 【table 2】fitting results with Lorentz model (confirmed 1.17)
\begin{table}[b]
	\centering
	\caption{Parameters $\tilde{\omega}_j/2\pi,\, \omega_j/2\pi$, and $\gamma_j/2\pi$ (THz) obtained by the fitting of experimental data to the Lorentz model. }
	\begin{tabular}{cllll}\hline
	\multicolumn{1}{c}{\ } &\multicolumn{1}{c}{~~$j=1$~~}&\multicolumn{1}{c}{~~$j=2$~~}&\multicolumn{1}{c}{~~$j=3$~~}&\multicolumn{1}{c}{~~$j=4$}\\
	\hline\hline
	$~\tilde{\omega}_j/2\pi$~~&~~$ ~13.8 $ &~~$ ~7.6$ &~~$ ~26.3 $~~ &~2514.6 \\
	$~\omega_j/2\pi$~~&~~$ ~13.5 $ &~~$~ 23.9 $ &~~$ ~32.0 $~~&~2546.5  \\
	$~\gamma_j/2\pi$~~&~~$ ~1.24$ &~~$~ 2.32$ &~~$ ~1.93$~~& ~340.6  \\\hline
	\end{tabular}
	\label{fit param}
\end{table}
%--------------------------------

%--------------------------------
%: 【table】 C_j
\begin{table}[b]
	\centering 
	\caption{Potential coefficients $C_3,\,C_4,\,C_6$, and $C_7$ in atomic units, where $R$ is the radius of the sphere in atomic units.}
	\begin{tabular}{ccccc}\hline
	\multicolumn{1}{c}{}&~~$C_3$&~~$C_4$&~~$C_6 (R)$&~~$C_7(R)$ \\\hline\hline
	Rb~~&~~0.733&~~$2.35 \times 10^3$ &~~$6.59 \times R^3$ &~~$3.87 \times 10^4 \times R^3$\\
	Cs~~&~~0.863&~~$2.95 \times 10^3$ &~~$7.77 \times R^3$ &~~$4.86 \times 10^4 \times R^3$ \\\hline
	\end{tabular}
	\label{C_i}
\end{table}
%--------------------------------
%--------------------------------
%: 【figure】 system
\begin{figure}[t]
\centering
%\aps{\includegraphics[width=8.5cm,pagebox=cropbox]{fig_3.pdf}}
\includegraphics[width=9cm]{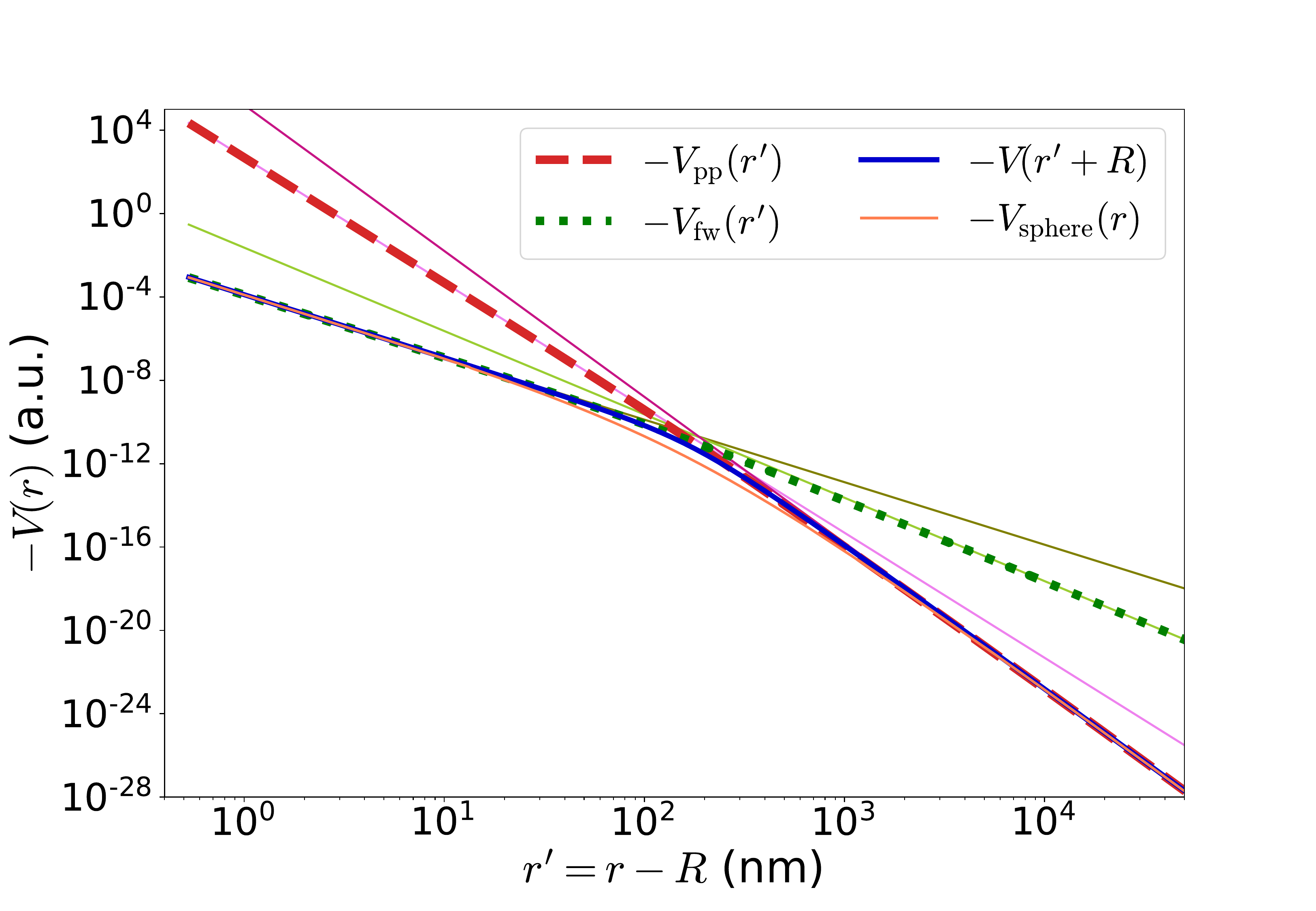}
\caption{Dispersion-force potential $V$ given by Eq.~(\ref{Vint}) with the shape function Eq.~(\ref{eq:shapef}) used throughout this paper, exact atom-nanosphere potential $V_{\rm sphere}$, atom-flat wall potential $V_{\rm fw}$, and atom-point particle potential $V_{\rm pp}$. Four thin lines denote $C_n/r^n$ for $n=3,4,6,7$ with the coefficients $C_n$ given by Table III. For all potentials, the common atomic polarizability and the spherical dielectric function are used. }
\label{fig:pot_compr}
\end{figure}
%--------------------------------

Figure~\ref{fit fig} shows the experimental data~\cite{Palik} and Eq.~\eqref{Lorentz eq} with the set of the optimal parameters. We notice that the Lorentz model is inaccurate near $\omega/(2\pi) \approx  3 \times 10^{6}$ GHz, which is due to the electronic absorptions consisting of the continuous bands~\cite{Fox}. However, this inaccuracy brings little detriment in the evaluation of $C_3$ and $C_6$ because of the following reasons. 
The atomic polarizability Eq. (\ref{one-oscillator}) sharply drops from unity to zero 
near the characteristic frequency $\omega_\mathrm{a}/(2\pi) \approx 0.4 \times 10^{6}$ GHz. 
Hence, the behavior of $\epsilon(\I \xi)$ in the region $\xi > \omega_\mathrm{a}$ does not contribute to the integrals, and it suffices to have a correct value of $\epsilon (\I \xi)$ only in the regime $\xi \le \omega_\mathrm{a}$ for the evaluations of the integrals for $C_3$ and $C_6$. 
The imaginary-frequency dielectric function $\epsilon (\I \xi)$ is related to the real-frequency one $\epsilon (\omega)$ via the formula \cite{LandauS}
\begin{align}
	\epsilon(\I \xi) = 1 + \frac{\pi}{2} \int_0^\infty \frac{\omega \epsilon''(\omega)}{\omega^2 + \xi^2} d \omega. 
\end{align}
This integral is dominated by the frequency in the regime $\omega \lesssim \xi$. Thus, $\epsilon(\omega)$ has an important contribution for $\omega \lesssim \omega_\mathrm{a}$, and the contribution from the dielectric function for $\omega > \omega_\mathrm{a}$ is negligible.

\subsubsection{Potential landscape}

By using the atomic polarizability Eq.~(\ref{one-oscillator}) and the dielectric function of silica glass Eq.~(\ref{Lorentz eq}) with constants tabulated in Tables \ref{valuestab} and \ref{fit param}, we numerically obtain the potential coefficients Eq.~(\ref{vdWcoefficientsSI}). These values in atomic units for rubidium and cesium atoms are tabulated in Table \ref{C_i}. 
The landscape of our potential between a cesium atom and a nanoparticle of radius $R=75$ nm is shown in Fig.~\ref{fig:pot_compr} where the exact potential $V_{\rm sphere}$~\cite{Buhmann}, the atom-flat wall potential $V_{\rm fw}$, and the atom-point particle potential $V_{\rm pp}$ evaluated with the common atomic polarizability and the dielectric function of the nanoparticle are also shown. Our potential reproduces the exact potential $V_{\rm sphere}$ quite well for any distances: both behave as $r'^{-3}$ at a short distance ($r' \lesssim 50$ nm) in agreement with the non-retarded behavior of $V_{\rm fw} (r')$, and as $r'^{-7}$ at a long distance ($r' \gtrsim 500 $nm) in agreement with the retarded behavior of $V_{\rm pp}$,  respectively, 
The middle distance (50 nm $\lesssim r' \lesssim $ 500 nm) corresponds to a crossover regime where potential is described by superpositions of the inverse powers of three, four, six, and seven. 
In general, the retardation effects of the electromagnetic field are important at distances $r' \gtrsim \lambda/(2\pi)$, where $\lambda/(2\pi)$ is in the range from 100 nm to 150 nm for cesium and rubidium atoms. 
Hence, the retardation is negligible as long as $V \propto r'^{-3}$, but is crucial at long distances where $V \propto r'^{-7}$. 
Except for a little difference between $V$ and $V_{\rm sphere}$ in the crossover regime, 
our potential agrees even quantitatively with the exact one. Since the most scattering phenomena are characterized by the long-range part of the potential, the use of the approximate potential $V$ does not change our results. 
Low-energy scattering by a cubic inverse potential~\cite{Muller, Alhaidari}, van der Waals potential~\cite{Idziaszek, Micheli, Julienne}, and an arbitrary single inverse power low potential~\cite{Cote96} are respectively studied. In the following sections, we investigate the scattering by our potential $V$ that renders characters of the several inverse powers depending on the distances, both classically and quantum mechanically.

%
% ___________________________________________________________________________________________
\section{Classical scattering} \label{sec:classical_scat}

%: classical absorption cross section

In the classical theory, the potential scattering is fully characterized by the atomic incident velocity $v = \sqrt{2E/\mu}$ 
and the impact parameter $\rho = L/(\mu v)$, where $E$ and $L$ are the incident energy and the continuous angular momentum, respectively. 
We consider a situation that a classical atom with velocity $v$ and impact parameter $\rho$ is incident on a nanosphere. 
Since the potential is a central field, the atom undergoes an effective potential, 
\begin{align}
V_{\rm eff}^{\rm (cl)} (r) = \frac{L^2}{2\mu r^2} + V(r), 
\end{align}
where the first term is the centrifugal potential, and $V(r)$ is the atom-surface interaction potential obtained in Sec.~\ref{interaction}. 
For a nonzero angular momentum, the effective potential has a centrifugal barrier in spite of the purely attractive nature of the bare potential $V(r)$. 
Since the centrifugal potential is proportional to $(\rho v)^2$, the barrier height of the potential $\underset{r}{\rm max} \{V_{\rm eff}^{\rm (cl)}  (r; \rho v)\} \equiv V^{\rm (cl)}_{\rm max} [\rho v]$ is a monotonous increasing function of $\rho v$. 
Whether the atom moves only in the outer region of the centrifugal barrier or enters the inner region of the barrier and is subsequently adsorbed at the surface due to the deep attractive potential, is also fully characterized by the product $\rho v$. 

% ___________________________________________________________________________________________
\subsection{Adsorption}

%--------------------------------
%: 【figure】classical capture range
\begin{figure}[b]
	\centering
	%\aps{\includegraphics[width=8.5cm,pagebox=cropbox]{fig_4.pdf}}
	\includegraphics[width=8.5cm]{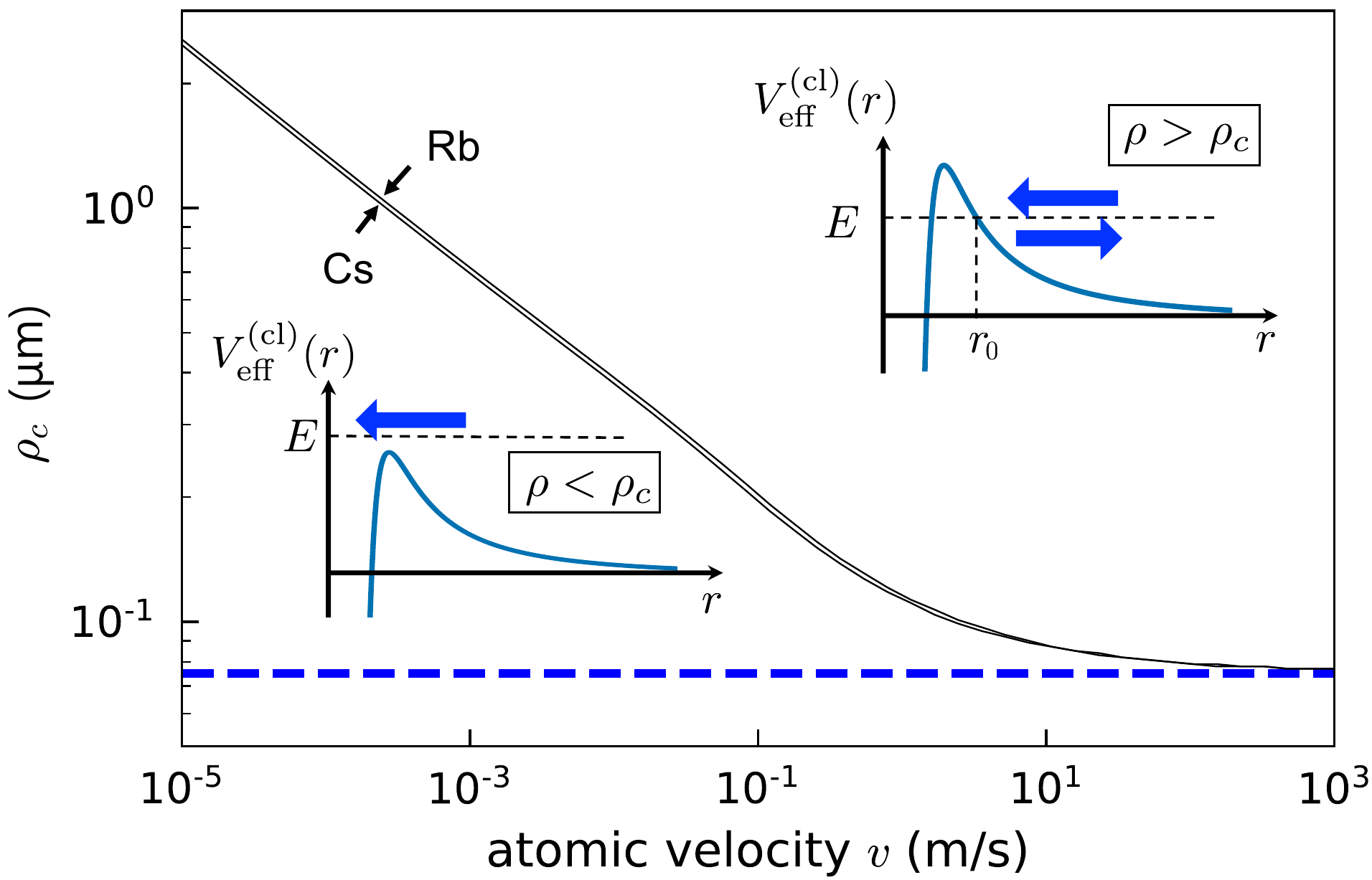}
	\caption{Classical capture range $\rho_c$ (solid curves) as a function of incident velocity $v$ for cesium and rubidium atoms. It behaves as $\rho_c \sim v^{-2/7}$ in the low-energy regime and approaches $R$ (horizontal dotted line) 
	in the high-energy limit. \Blue{The region of velocity displayed here roughly comparable to temperature regime 
	$O(10^{-12})$ K $\lesssim 2E/k_B \lesssim O(10^4)$ K}.}
	\label{fig:cl}
\end{figure}
%----------------------------------
The atom is adsorbed at the surface when the incident energy $E$ is larger than the barrier height $V_{\rm max}^{\rm (cl)} [\rho v]$. 
For each incident velocity $v$, we thus define a {\it classical capture range} $\rho_c$ by
\begin{align}
\frac{1}{2}\mu v^2 = V^{\rm (cl)}_{\rm max} [\rho_c v], \label{eq:cap}
\end{align}
so an incident atom with the impact parameter $\rho < \rho_c$ is adsorbed at the surface, while an atom with $\rho > \rho_c$ is not. The quantity $\pi \rho_c^2$ corresponds to the {\it classical absorption cross section}~\cite{LandauM, Friedrich2013}.

Figure~\ref{fig:cl} shows the classical capture range $\rho_c$ as a function of the incident velocity $v$. We find that the capture range in the experimentally achievable low-energy regime is more than ten times larger than the geometric radius of the nanosphere. The capture range approximately behaves as
\begin{align}
\rho_c = R + \zeta_n v^{-2/n},\quad \zeta_n = \frac{\sqrt{n}}{(n-2)^{\frac{n-2}{2n}}}\left(\frac{C_n}{\mu}\right)^{1/n}, \label{eq:rhoc}
\end{align}
where the second term is the classical capture range for a single inverse power-law potential $- C_n/r^n~(n > 2)$ between two point particles. The power $n$ corresponds to the long-range behavior of our potential, i.e., $n=7$. 
For slow atoms, the second term in Eq.~(\ref{eq:rhoc}) is dominant, and $\rho_c$ thus behaves as $\sim v^{-2/7}$. As $v$ increases, it starts to deviate from 
$v^{-2/7}$ around $v \sim$ a few cm/s, at which the capture range measured from the surface $\rho_c - R$ decreases to a few hundreds nanometer, since the potential starts to deviate from $r'^{-7}$, entering a crossover regime around this distance. In a high-energy limit, on the other hand, the capture range coincides with the geometric radius of the sphere, $\rho_c \to R$. In this limit we may regard the potential as an inverted hard-wall potential of radius $R$. 

The results so far also hold in the quantum scattering theory, as we see in later sections. 
However, we note that in the classical theory the atoms of angular momentum $L=0$ are always captured at the surface for any incident velocity because of the absence of the centrifugal barrier. This is {\it not} the case in the quantum theory. 
% ___________________________________________________________________________________________
\subsection{Elastic differential cross section}
%--------------------------------
%: 【figure】classical loci and differential cross sections
\begin{figure}[b]
	\centering
	%\aps{\includegraphics[width=8.3cm,pagebox=cropbox]{fig_5.pdf}}
	\includegraphics[width=8.3cm]{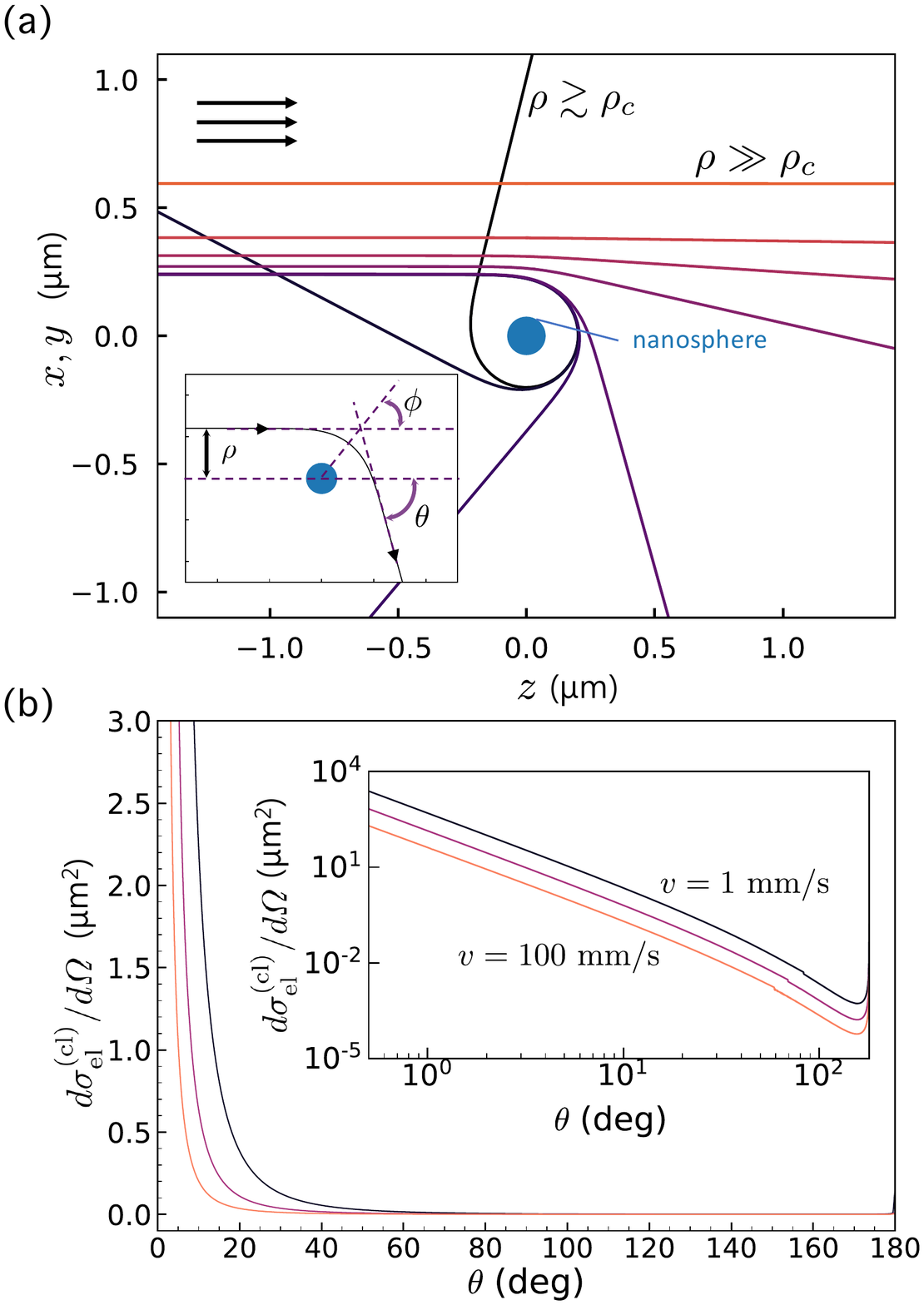}
	\caption{(a) Classical scattering trajectories for various impact parameters. The atom is incident from left to right along the $z$ axis with the velocity $v = 50$~mm/s, and the corresponding capture range is $\rho_c \approx 239$~nm. Trajectories for $\rho-\rho_c = 6.5 \times 10^{-4},~ 1.1 \times 10^{-2},~ 0.2,~ 1.8,~32,~74,~144$, and $356~$nm are drawn. The atom is largely deflected for $\rho \gtrsim \rho_c$ while it is little affected by the potential for $\rho \gg \rho_c$. Inset schematically shows the scattering angle $\theta$ for $\rho - \rho_c = 1.8~$nm. 
	(b) Classical differential cross sections $d\sigma_{\rm el}^{\rm (cl)}/d\varOmega$ for incident velocities $v = 1,~10,~100$ mm/s. 
	The angular distribution is narrower for larger incident velocities. Inset shows the same quantities in log scale. 
	}
	\label{fig:cl_elas}
\end{figure}
%----------------------------------

% --- trajectories

We show in the previous subsection that for a fixed energy, an atom of impact parameter $\rho$ smaller than $\rho_c$ is adsorbed onto the surface of the nanosphere and cannot be detected. For $\rho > \rho_c$, on the other hand, the motion of an atom is restricted in the outer region of the centrifugal barrier $r > r_0$, where $r_0$ is the classical turning point defined as $E = V^{\rm (cl)}_{\rm eff} (r_0)$ (see, inset of Fig.~\ref{fig:cl}). 
Figure~\ref{fig:cl_elas} (a) shows the atomic trajectories for several impact parameters $\rho > \rho_c$ 
when an atom with the velocity $v = 50~$mm/s is injected along the $z$ axis from minus infinity to the positive $z$ direction, subsequently deflected due to the atom-surface potential, and scattered to an angle $\theta$ asymptotically. 
The trajectory of the atomic motion is obtained by integrating the equation of motion in the relative polar coordinates, 
\begin{align}
\dot{r} = v \sqrt{1- \frac{\rho^2}{r^2} - \frac{2V (r)}{m v^2}},\quad \dot{\phi} = \frac{\rho v}{r^2}, 
\end{align}
with the initial condition $ r = \infty, ~\phi = \pi$. When the impact parameter is precisely equal to $\rho_c$, the atom eternally orbits around the nanosphere with radius $\rho_c$. For impact parameters slightly larger than $\rho_c$, the atom is largely deflected by the potential, orbiting around the nanosphere for a while, and eventually scattered to a certain angle $\theta$. For larger impact parameters, in contrast, the motion of the atom is less and less affected by the potential and the scattering angle $\theta$ is thus smaller. 

% --- differential cross section

In a typical scattering experiment, not a single atom but a beam consisting of many atoms with a certain velocity and various impact parameters is incident on a target, and a detector located at a solid angle $d\varOmega = 2 \pi \sin \theta d\theta$ sufficiently far from the scattering center, counts the number of the scattered atoms. 
The detector counts located at a solid angle $d\varOmega$ per unit time yield the differential cross section, 
\begin{align}
\frac{d\sigma_{\rm el}^{\rm (cl)}}{d\varOmega} = \frac{\rho (\theta)}{ \sin \theta} \left|\frac{d\rho}{d\theta}\right|\qquad (\rho >\rho_c), \label{eq:cl_diff}
\end{align}
where we define the scattering angle within $0 \le \theta \le \pi$, by summing up all the branches associated with the multi-valuedness of the impact parameter $\rho(\theta)$. 
In Fig.~\ref{fig:cl_elas} (b), we show the angle dependence of the differential cross section for several velocities of the atomic beam. For a faster beam, the angular distribution is narrower and the forward scattering is more dominant. The sharp increase in the differential cross section near $\theta = \pi$ is the analogy of the glory appearing in the Brocken effect~\cite{Friedrich2013}.

The classical differential cross section diverges as $\theta \to 0$ and hence the total elastic cross section obtained by integrating with respect to the solid angle also diverges for any incident velocity. 
This unphysical divergence arises because the infinite series of impact parameters are involved in the classical differential cross section Eq.~(\ref{eq:cl_diff}) by definition, and all of these atoms are scattered to an infinitesimally small angle due to the potential tail no matter how large the impact parameter is. Realistically, the beam width is finite and thus the divergences of the forward scattering and total elastic cross section do not occur.

% ___________________________________________________________________________________________
\section{Quantum theory of scattering} \label{st}

In this section we investigate the potential scattering in the quantum theory. 
As the temperature is lowered, the thermal de Broglie wavelength becomes comparable with or longer than the size of the nanoparticle. We study the quantum effects in such a low-energy regime. 

% ___________________________________________________________________________________________
\subsection{Formulation}

We consider the stationary state described by the Schr\"odinger equation for the relative motion, 
\begin{align}
\left[-\frac{\hbar^2 \nabla^2}{2\mu} + V (\bm{r})\right] \psi (\bm{r}; k) = E  \psi (\bm{r}; k), 
\end{align}
when the plane wave of the wavenumber $k=\mu v/\hbar$ and the energy $E = \hbar^2 k^2 /(2\mu)$ is incident on the nanosphere, as shown in Fig.~\ref{fig:system}. 
The stationary state $\psi(\bm{r}; k)$ is expanded in terms of the partial waves labeled by the angular-momentum quantum number $l$, and its asymptotic form is given by 
\begin{align} 
	\psi(\bm{r}; k) &= \sum_{l=0}^\infty \frac{u_l (r;k)}{r} P_l(\cos \theta)\\
	&\xrightarrow{r \to \infty} e^{\I k z} + \frac{f(\theta; k)}{r} e^{\I k r},\label{eq:psi_asym}
\end{align}
where $u_l (r; k)$ denotes the $l$th radial function, $P_l(x)$ the Legendre functions, 
$\theta$ the scattering angle, and $f(\theta; k)$ the scattering amplitude, respectively. 

% ----- radial Sch. eq

The problem is now reduced to solve the equation of the radial function for each partial wave:  
\begin{align}
\frac{d^2 u_l (r; k)}{d r^2} = - k_l(r; k)^2 u_l(r; k), \label{Sch-eq}
\end{align}
where $k_l(r;k)$ is the local wave number of the $l$th partial wave defined as \cite{Friedrich2013}
\begin{align}
	k_l(r; k) = \sqrt{k^2 - \frac{l(l+1)}{r^2} - \frac{2\mu V(r)}{\hbar^2}}. \label{LWN}
\end{align}
Equation~\eqref{Sch-eq} is equivalent to the one-dimensional Schr\"odinger equation with 
the effective potential 
\begin{align}
V_{\rm eff}^{(l)} (r) = \frac{\hbar^2 l (l+1)}{2\mu r^2} + V(r), 
\end{align}
where the first term corresponds to the centrifugal potential for the $l$th partial wave, and the second term is the dispersion-force potential obtained in Sec.~\ref{interaction}.

%: --- large distance 

At large distances from the surface, the asymptotic form of the $l$th radial wavefunction is expressed in terms of the $l$th order spherical Bessel function $j_l(x)$ and the $l$th order spherical Neumann function $n_l(x)$ as
\begin{align} 
	u_l(r; k) &\xrightarrow{r \to \infty} k r \left[ F_l(k) j_l(k r) - G_l(k) n_l(k r) \right]  \label{AsF}, 
\end{align}
which is also expressed in terms of the diagonal elements $S_l(k)$ of the $S$ matrix, 
\begin{align}
	u_l(r; k)\xrightarrow{r \to \infty} \frac{(2l+1)(-1)^{l+1}}{2\I k} \left[e^{-\I k r} - (-1)^l S_l(k) e^{\I k r}\right]. \label{Asym_Sl}
\end{align}
The diagonal elements $S_l (k)$ of the $S$ matrix are expressed in terms of the phase shift $\delta_l$ as $S_l(k) = e^{2\I \delta_l(k)}$. 
The asymptotic forms of the spherical Bessel functions $j_l(x) \xrightarrow{x \to \infty} \sin\left(x - l \pi/2 \right)/x$ and the spherical Neumann functions $n_l(x) \xrightarrow{x \to \infty} - \cos\left(x - l \pi/2 \right)/x$ yield the relation
\begin{align}
	S_l(k) = \frac{F_l(k) + \I G_l(k)}{F_l(k) - \I G_l(k)}.\label{SFG}
\end{align}
If the $l$th partial wave is unaffected by the potential, the phase shift is zero and hence $S_l (k) =1$. The effects of the potential scattering on the $l$th partial wave are characterized by the deviation of the value $S_l (k)$ from unity. 

If the incident wave is partially absorbed by the surface, which is the case we consider, 
the $S$ matrix is non-unitary $|S_l(k)| < 1$, and the phase shift is complex~\cite{LandauQ}. As we discuss in the next subsection, the reflection of the $l$th partial wave can occur at a nonclassical region in the coordinate space, 
and the only portions that are transmitted through the nonclassical region are absorbed by the surface. The absorption probability of the $l$th partial wave is given by $1-|S_l (k)|^2$. 

Scattering amplitude $f(\theta; k)$ in Eq.~\eqref{eq:psi_asym}, which is 
written in terms of the $S$-matrix as 
\begin{align}
	 f(\theta; k) &= \sum_{l=0}^\infty \frac{(2l+1)}{2 \I k} (S_l(k) - 1) P_l(\cos \theta) \label{ScattA}, 
\end{align}
characterizes the angle dependence of the scattering. 
The elastic differential cross section is defined in terms of the scattering amplitude as 
\begin{align}
	&\frac{d\sigma_\mathrm{el}}{d \varOmega}(\theta; k) = |f(\theta; k)|^2, \label{eq:q-diff-scat}
\end{align}
which involves interference between different $l$th partial waves. 
Summing over the entire solid angle eliminates the off-diagonal terms, and yields the total elastic cross section, 
\begin{align}
\sigma_\mathrm{el}(k) = \sum_{l=0}^\infty \frac{(2l+1) \pi}{k^2} |S_l(k) -1|^2 = \sum_{l=0}^\infty \sigma_\mathrm{el}^{(l)}(k). \label{eq:sigma_el}
\end{align}
The scattering amplitude and the elastic cross section are defined solely by the elastically scattered wave under the influence of the potential, since the contribution of the unaffected wave is eliminated by subtracting 1 from $S_l$. 
On the other hand, the absorption cross section $\sigma_\mathrm{abs}(k)$ is written \cite{LandauQ} as 
\begin{align}
\sigma_\mathrm{abs}(k) = \sum_{l=0}^\infty \frac{(2l+1) \pi}{k^2} (1 - |S_l(k)|^2) = \sum_{l=0}^\infty \sigma_\mathrm{abs}^{(l)}(k). \label{crossection_eq}
\end{align}
The elastic and absorption cross sections involve only diagonal terms of the $S$ matrix and satisfy the optical theorem:
\begin{align}
\sigma_{\rm el}(k) + \sigma_{\rm abs} (k) = \frac{4\pi}{k} {\rm Im}[f(\theta=0; k)]. \label{optical_theorem_eq}
\end{align}

% ___________________________________________________________________________________________
\subsection{Boundary condition near the surface}

The atom-surface interaction is strongly attractive near the surface of the nanosphere, thereby the waves are destined to be absorbed once they approach very close to the surface. 
There are several ways to take the absorption into account. One of methods is to make the potential complex as introduced in the study of the reactive collisions of molecules~\cite{Idziaszek}. However, the atom-surface potential is nontrivial in the proximity to surface and it is thus too obscure to determine a concrete form of the potential. 
Alternatively, we impose the following boundary condition in the vicinity of the surface $r\to R+$~\cite{Effective-range, Friedrich2013, Cote96, Cote97, Cote98}: 
\begin{align}
	u_l(r; k) \overset{r \to R+}{\propto} \frac{1}{\sqrt{k_l(r; k)}} \exp \left( - \I \int^r k_l(\rho; k) d \rho \right). \label{WBC}
\end{align} 
With this boundary condition, in other words, we make an ansatz so there is only an incoming wave and no outgoing wave in the vicinity of the surface. 
The form of the wavefunction Eq.~\eqref{WBC} is based on the semiclassical Wentzel-Kramer-Brillouin (WKB) approximation, which is shown to be valid near the surface $r \to R+$ for the following reasons. 

%--------------------------------
%: 【figure】quantality function  
\begin{figure}[b]
\centering
%\aps{\includegraphics[width=8.5cm,pagebox=cropbox]{fig_6.pdf}}
\includegraphics[width=8.5cm]{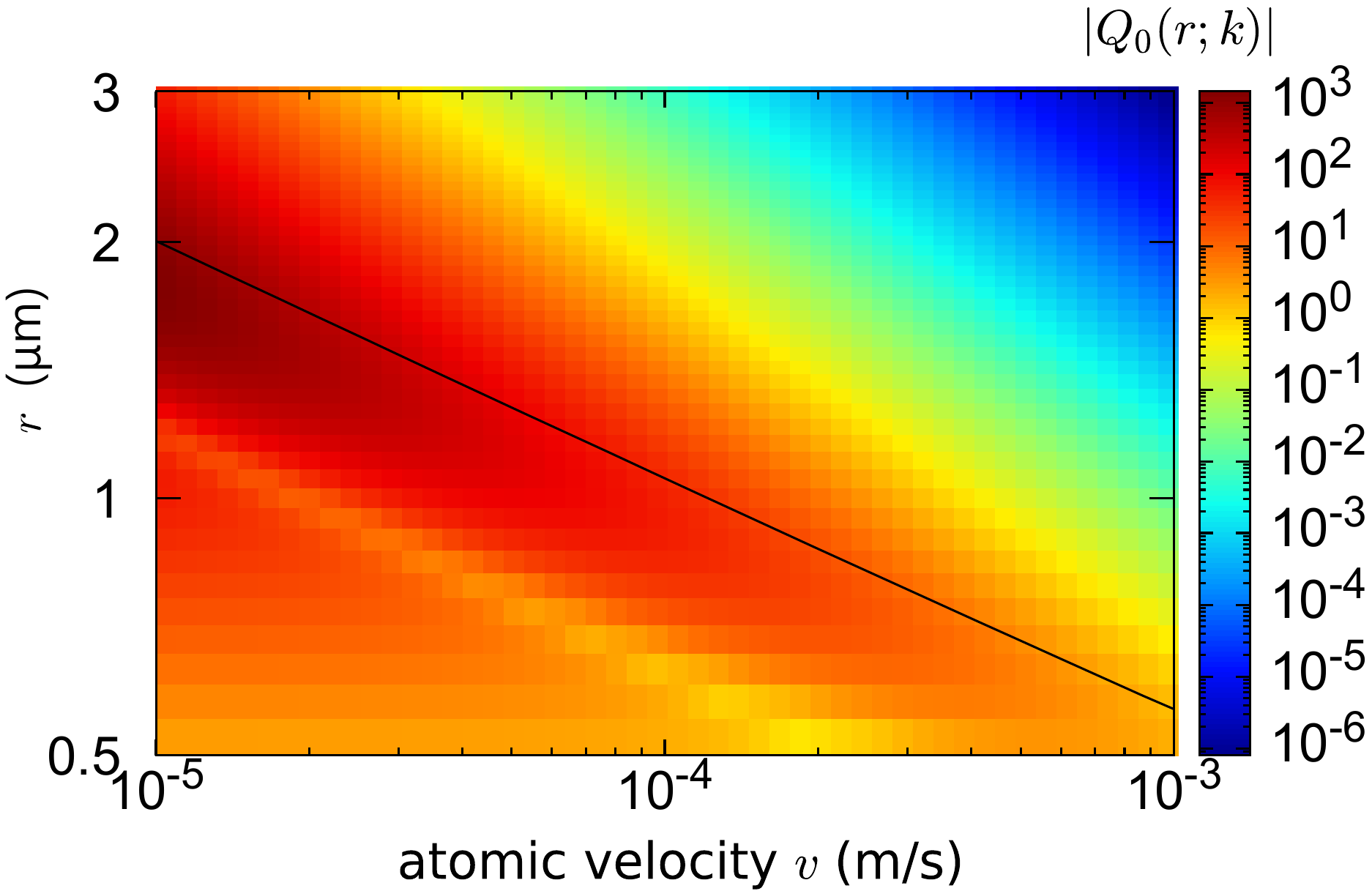}
\caption{The magnitude of the $s$-wave quantality function $|Q_0 (r;k)|$ versus incident velocity of a rubidium atom. Solid line shows the distance $r \propto v^{-2/7}$ at which $|Q_0 (r;k)|$ takes the maximal value for each $v$.}
\label{fig_Q}
\end{figure}
%----------------------------------

Here we revisit the essence of this boundary condition by referring the classical capture model. 
In the general WKB approximation, the exponential form of the wavefunction is retained but the exponent is replaced by the action integral as 
\begin{align} 
u_{\rm WKB} (r; k) \propto  \exp\left( \pm \I  \int^r  k_l (\rho ; k) d\rho \right).
%u_{\rm WKB} (r; k) \propto A(r; k) \exp\left( \pm \I  \int^r  k_l (\rho ; k) d\rho \right).  
\end{align}
This wavefunction is valid, or can be even exact, as long as the {\it quantality function} or {\it badlands function}
\begin{align}
Q_l(r; k) = k_l(r;k)^{-3/2} \frac{d^2}{d r^2} k_l(r;k)^{-1/2}, 
\end{align}
satisfies $|Q_l(r; k)| \ll 1$. The quantality function $Q_l(r; k)$ is thus a measure of nonclassicality~\cite{Friedrich2002}. 
In the regime where $|Q_l(r; k)| \ll 1$ is satisfied, the incoming wave $\propto \exp(-\I \int^r k_l (\rho; k)d\rho)$ and outgoing wave $\propto \exp(+\I\int^r k_l (\rho; k)d\rho)$ are unambiguously decomposed.

Now we go back to our specific problem. In the vicinity of the surface, the interaction potential dominates in Eq.~(\ref{LWN}), i.e., 
$k_l(r;k) \xrightarrow{r \to R} \sqrt{- 2\mu V(r)/\hbar^2}$. 
For the potential obtained in Sec.~\ref{interaction}, the local wavenumber behaves as 
$k_l(r;k) \xrightarrow{r \to R}  \sqrt{\beta_3}/(r-R)^{3/2}$, and 
the quantality function thus behaves as $Q_l(r;k) \overset{r \to R}{\propto} (r - R)/\beta_3$, which indicates that 
the WKB approximation is valid at small distances.  
At sufficiently large distances where the atom is essentially free from the interaction and 
the local wave number $k_l(r;k)$ behaves as $k$, the condition $|Q_l(r; k)| \ll 1$ is also satisfied. 
Thus the exact wavefunction for the $l$th partial wave is well described by the superposition of ingoing and outgoing waves $\propto \exp(\pm \I\int^r k_l (\rho; k)d\rho)$ at large distances. 
At small distances from the surface, in contrast, it would be described by only incoming wave $\propto \exp(-\I \int^r k_l (\rho; k)d\rho)$ because the wave cannot go outward due to the strong attractive interaction. 
This ansatz implies that the reflection of the $l$th incoming wave occurs somewhere in the intermediate distances at which $|Q_l (r; k)| \gg 1$.

For $l \ge 1$, we can estimate a critical angular momentum $\hbar l_c$ for each velocity from the condition $\hbar^2 k^2/(2\mu) \simeq \underset{r}{\rm max}\{V_{\rm eff}^{(l_c)} (r)\}$, which is the quantum version of Eq.~(\ref{eq:cap}), such that the $l$th partial waves 
are absorbed if a quantum analog of the impact parameter $\hbar l / (\mu v) = l/k$ 
is smaller than $l_c/k$, while they are elastically scattered by the centrifugal barrier or unaffected by the potential if $l/k > l_c/k$. The quantality function for the $l$th partial wave $Q_{l> l_c} (r; k)$ diverges when $k_{l >l_c} (r_0; k)= 0$, thus we may infer that the elastic scattering occurs around $r_0$ as previously discussed in this subsection. We note that the equation $k_l (r_0; k)= 0$ is the quantum version of the condition that determines the classical turning point $E=V_{\rm eff}^{\rm (cl)} (r_0)$. The elastic scattering for $l > l_c$ is thus regarded as a classically allowed reflection by the centrifugal barrier. 

The partial wave of $l=0$ is of particular interest. Classically, atoms with zero angular momentum are totally adsorbed onto the surface for any incident velocity because of the absence of the centrifugal barrier. 
Quantum mechanically, on the other hand, $s$-wave can be reflected even though the potential is purely attractive. This is a classically forbidden reflection, namely, the {\it quantum reflection}~\cite{Friedrich2013}. The quantum reflection is expected to occur in a coordinate space where $|Q_0 (r;k)| \gg 1$ as if an effective mirror exists in there. Such a nonclassical spacial region is called {\it badlands}~\cite{Friedrich2002, Segev, Effective-range, Friedrich2013, Cote96, Cote97, Cote98}.

Figure~\ref{fig_Q} shows the magnitude of the $s$-wave quantality function for a rubidium atom. 
For lower incident energies, the magnitude of the $s$-wave quantality function as well as the distance from the surface at which $|Q_0 (r; k)|$ 
takes the maximal value are larger. 
This indicates that the position of the badlands, i.e., an effective mirror, moves outer and its reflectivity is larger for lower incident energies. 
The location of the effective mirror, defined here as the position $r$ at which $|Q_0 (r;k)|$ takes the maximal value (a solid line in Fig.~\ref{fig_Q}), behaves as $r \propto v^{-2/7}$ at low energies. 
For higher energies, the peak height of $|Q_0 (r; k)|$ decreases, and is indiscernible in the high-energy limit. 

Quantum reflection has been experimentally observed and studied in several systems. Experiments on quantum reflection on fluid surfaces have been carried out by measurements of the reflectivity or sticking probability of incident helium or hydrogen atoms scattered by a liquid helium surface~\cite{Nayak, Berkhout, Doyle, Yu}. Quantum reflection on solid surfaces has also been observed: specular reflection of cold metastable neon atoms on a silicon and a BK7 glass surface~\cite{Shimizu2001, Shimizu2002},  quantum reflection of helium atoms incident on a silicon surface~\cite{Oberst}, of Bose-Einstein condensates on a solid surfaces~\cite{Pasquini2004, Pasquini2006}, and far from threshold~\cite{Druzhinina}.

% ___________________________________________________________________________________________
\section{Results}\label{results}  

We numerically solve the radial Schr\"odinger Eq. (\ref{Sch-eq}) for each partial wave $l$ with boundary condition Eq.~(\ref{WBC}) at a point $r=R+$.  
The obtained asymptotic values $F_l(k)$ and $G_l(k)$ then yield the diagonal elements $S_l(k)$ of the $S$ matrix according to Eq.~(\ref{SFG}). 
In this section, we first study an extremely low-energy regime where the $s$-wave scattering is dominant. In later subsections, we investigate the dependencies of the scattering properties on the atomic incident velocity and on the radius of the nanoparticle. 

% ___________________________________________________________________________________________
\subsection{$s$-wave scattering}

In the previous section, we denoted that $s$-wave scattering is qualitatively different from the classical scattering of atoms with vanishing angular momentum. 
In this subsection, we show various $s$-wave scattering properties, including the scattering length, the differential cross section, the elastic, and the absorption cross sections in a sufficiently low-energy regime.

%:  s-wave

%--------------------------------
%: 【figure】 potential, badlands function, wavefunction
\begin{figure}[t]
\centering
%\aps{\includegraphics[width=8.7cm,pagebox=cropbox]{fig_7.pdf}}
\includegraphics[width=8.7cm]{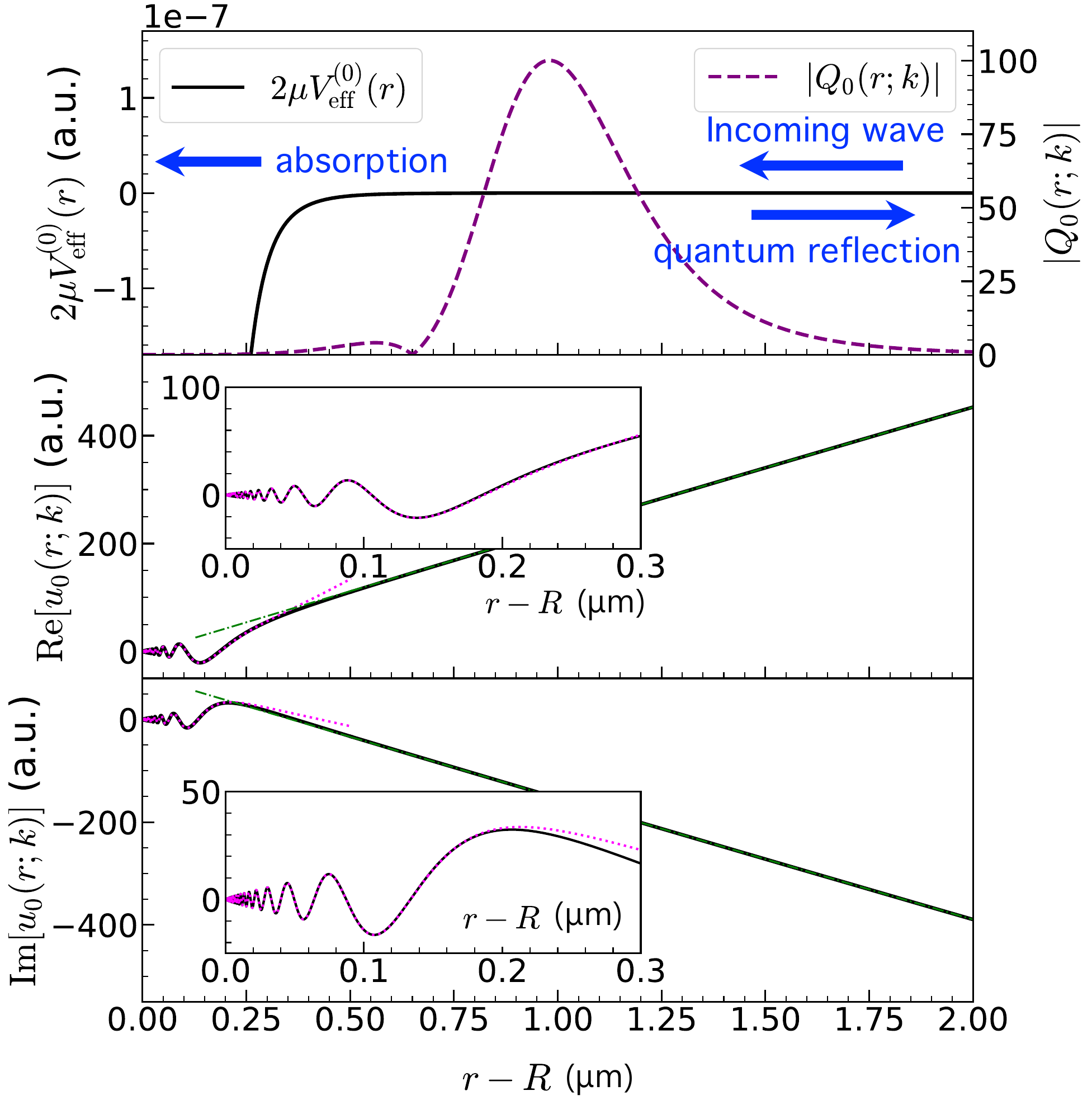}
\caption{Top panel shows the $s$-wave effective potential $2\mu V_{\rm eff}^{(0)}(r)=2 \mu V(r)$ (left reference) and the magnitude of the quantality function $|Q_0(r; k)|$ (right reference) for $R=75$ nm and a rubidium atom of the incident velocity $v= 100$ {\textmu}m/s. Middle and bottom panels show the real and imaginary parts of the $s$-wave radial wavefunction, respectively. Solid curves are the exact solution of the Schr\"odinger equation with the WKB boundary condition, dotted curves are the semiclassical WKB wavefunction, and the dashed curves are the asymptotic form given by Eq.~\eqref{AsF}. The insets enlarge the domain of the atom-surface distance smaller than 300 nm. }
\label{fig:pot}
\end{figure}
%--------------------------------

The top panel of Fig.~\ref{fig:pot} shows the landscapes of the interaction potential in atomic units, as well as the $s$-wave quantality function for the incoming atomic velocity $v = 100$~\textmu m/s (see also Fig.~\ref{fig_Q}). 
In the vicinity of surface $r-R \to 0$, and at large distances $r \to \infty$, 
the magnitude of the quantality function $Q_0 (r; k)$ is small but is significantly large at intermediate distances. 
As discussed in the previous section, a portion of the wave approaching the surface through the badlands region is 
lost due to the absorption, while the remaining portion that does not go through the badlands undergoes {\it quantum} reflection. 
In the middle and bottom panels of Fig.~\ref{fig:pot}, we show the radial function $u_0 (r; k)$ (solid curve), its asymptotic form~Eq.~\eqref{AsF} away from the surface (dashed-dotted curve), and the WKB wavefunction that has only an incoming wave (dotted curve). We find that the exact wavefunction is well described by the WKB wavefunction from the surface proximity up to an appreciable distance ($\approx$ 200 nm) from the surface. 
The badlands region is located at a considerably large distance ($\approx 1$ \textmu m) from the surface, where the potential is weakly attractive and the wavefunction is described by the asymptotic form.

%:  zero-energy limit

%--------------------------------
%: 【table】k=0
\begin{table}[b]
	\centering 
	\caption{Zero-energy limit of the $s$-wave scattering length, the differential elastic cross section, and the elastic cross section for a fixed radius of nanosphere ($R=75$ nm). }
	\begin{tabular}{ccc}\hline
	\multicolumn{1}{c}{} & Cs & Rb \\
	\hline\hline
	$A_0$ & $~~(0.30 - \I \, 0.19)$~\textmu m & $~~(0.27 - \I \, 0.17)$~\textmu m\\
	$d\sigma_{\rm el}(0)/d\varOmega$& $0.13$~\textmu m$^2$  &$0.10$~\textmu m$^2$ \\
	$\sigma_{\rm el}(0)$ & $1.60$~\textmu m$^2$ & $1.27$~\textmu m$^2$\\
	\hline
	\end{tabular}
	\label{tab_k0}
\end{table}
%--------------------------------

In the limit $k \to 0$, there is no scattering of the partial waves for $l\ge 1$, i.e., $S_{l \ge 1}(k) \approx 1$, and hence the $s$-wave contribution dominates all scattering properties. 
In terms of the complex $s$-wave scattering length $A_0$~\cite{Friedrich2013}, the asymptotic behavior of 
the $s$-wave wavefunction and the phase shift at $k \to 0$ are given by $u_0(r;0) \overset{r \to \infty}{\propto} r - A_0$, and $\delta_0(k) \xrightarrow{k \to 0} - k A_0$, respectively. 
We thus obtain the dominant element of the $S$ matrix in the extremely low-energy regime as 
\begin{align}
S_0(k) \xrightarrow{k \to 0} e^{2k \mathrm{Im}[A_0]} e^{-2\I k \mathrm{Re}[A_0]}\label{S0threshold}. 
\end{align}
The elastic differential cross section,  the elastic cross section associated with the quantum reflection, and the absorption cross section in the zero-energy limit are written as~\cite{Friedrich2013, LandauQ}
\begin{align}
	&\frac{d\sigma_\mathrm{el}}{d \varOmega}(\theta; k) \xrightarrow{k \to 0} |A_0|^2, \quad 
	\sigma_\mathrm{el}(k) \xrightarrow{k \to 0} 4 \pi |A_0|^2, \label{eldiff0} \\
	%\quad v\sigma_{\mathrm{el}}(k) \xrightarrow{k \to 0} \frac{4\pi \hbar k}{\mu}|A_0|^2 \label{el0} \\
	&\sigma_\mathrm{abs}(k) \xrightarrow{k \to 0} -\frac{4\pi}{k} \mathrm{Im}[ A_0 ], 
	%v \sigma_{\rm abs}(k) \xrightarrow{k \to 0} -\frac{4\pi \hbar}{\mu} {\rm Im}[A_0]\label{abs0}. 
\end{align}
in consistent with the Wigner threshold law~\cite{Wigner}. 
The numerically obtained zero-energy wave function yields the value of $A_0$, and the results are summarized in Table~\ref{tab_k0}.

% ___________________________________________________________________________________________
\subsection{Incident velocity dependence}\label{T-dep}

We next investigate various cross sections when the incident velocity is varied while the radius of the sphere is fixed as $R=75$~nm. In the following we show results for cesium atoms unless otherwise stated, since we have qualitatively the same results for rubidium atoms. 

%--------------------------------
%: 【figure】 elastic diff cross sections vs v
\begin{figure}[b]
\centering
%\aps{\includegraphics[width=8.7cm,pagebox=cropbox]{fig_8.pdf}}
\includegraphics[width=8.7cm]{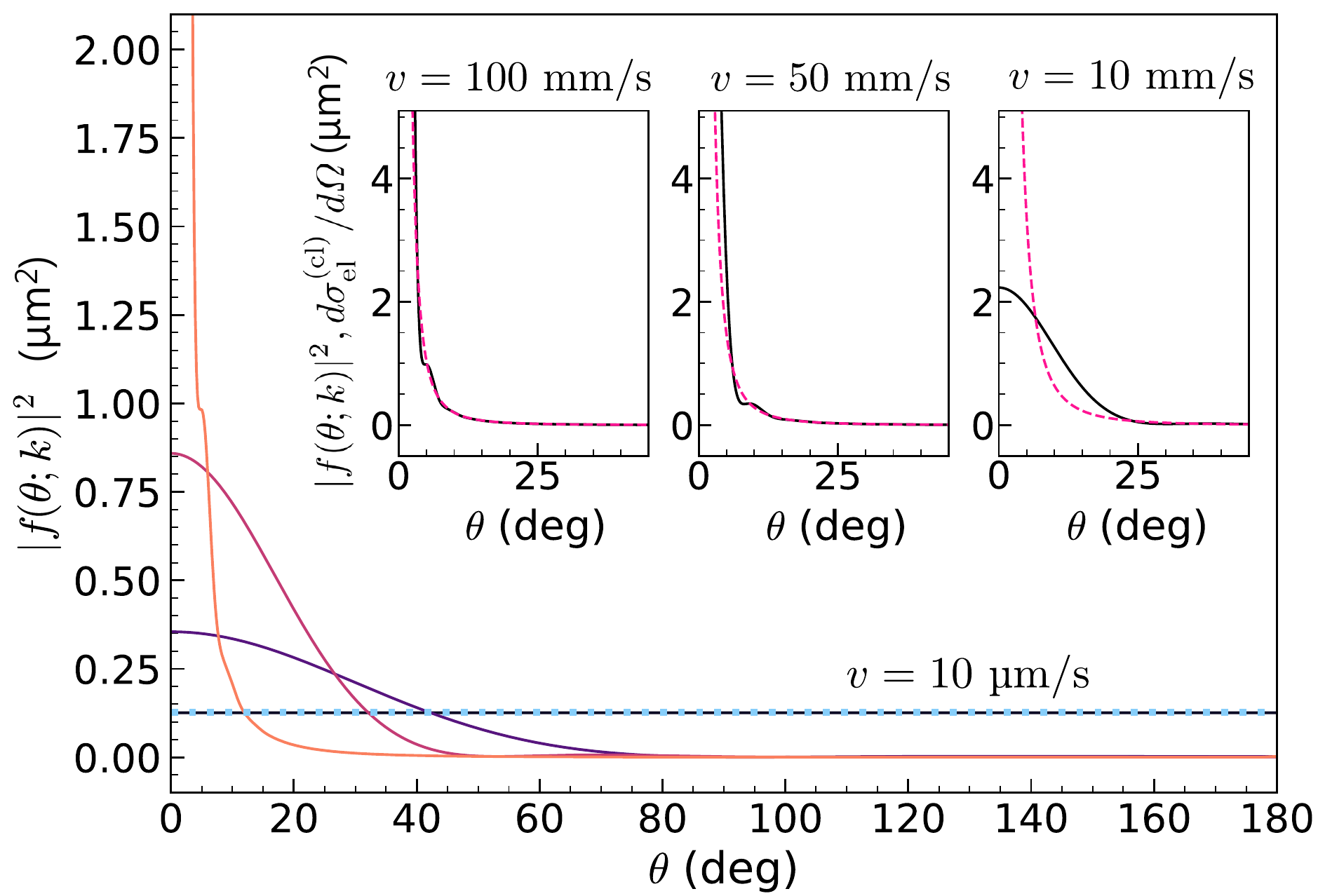}
\caption{Elastic differential cross sections for cesium atoms of the incident velocities $v =100,~5,~2.5~$mm/s, and $10$~\textmu m/s. The anisotropy of the scattering is suppressed as the energy decreases, and $|f(\theta; k)|^2$ is independent of the angle, approaching a constant value $|A_0|^2$ (horizontal dotted line) in the low-energy limit. 
Insets compare the quantum (solid curve) and classical (dotted curve) differential cross sections for fixed velocities. }
\label{DECS}
\end{figure}
%----------------------------------
%-------------------------------------
%: 【figure】cross sections (Cs)
\begin{figure}[h]
\centering
%\aps{\includegraphics[width=8.5cm,pagebox=cropbox]{fig_9.pdf}}
\includegraphics[width=8.5cm]{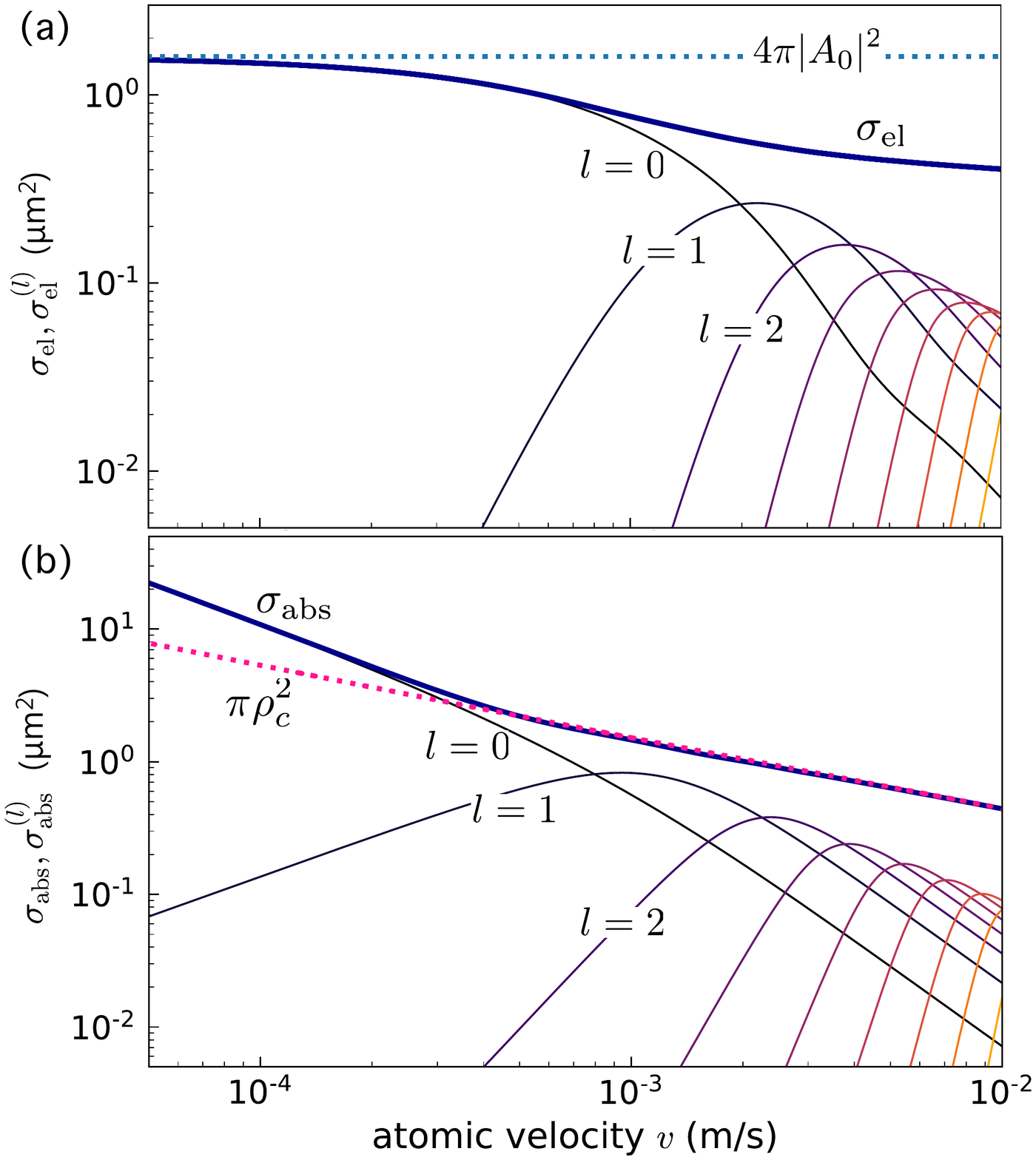}
\caption{Scattering cross sections for cesium atoms. 
 (a) Total elastic cross section $\sigma_{\rm el}$ (thick solid curve), and partial elastic cross sections $\sigma_{\rm el}^{(l)}$ for the low-lying partial waves (thin solid curves). In the low-energy limit, $\sigma_{\rm el}$ approaches $4\pi |A_0|^2$ (horizontal dotted line). 
 (b) Total absorption cross section $\sigma_{\rm abs}$ (thick solid curve), partial absorption cross sections $\sigma_{\rm abs}^{(l)}$ (thin solid curves), and classical absorption cross section $\pi \rho_c^2$ (dotted line) with $\rho_c$ being defined by Eq. \eqref{eq:cap}. }
\label{fig_sigma_Cs}
\end{figure}
%--------------------------------------

%: differential cross section

Figure~\ref{DECS} shows the elastic differential cross sections $|f(\theta; k)|^2$ for several incident velocities. 
For faster atoms, the forward scattering is more dominant and the angular distribution of $|f(\theta; k)|^2$ is narrower, in a manner similar to the classical differential cross section~(see also Fig.~\ref{fig:cl_elas}). 
At large velocities, the angular distributions of the quantum and classical differential cross sections agree well, as shown in 
the leftmost inset of Fig.~\ref{DECS}. 
As the velocity decreases, the anisotropy of $|f(\theta; k)|^2$ is suppressed, and in the $s$-wave regime it is independent of angle, approaching a constant value $|A_0|^2$. 
The difference between the angular distributions of the quantum and classical differential cross sections starts to be evident around $v \approx O(10)$ mm/s,  where the thermal de Broglie wavelength is comparable to the size of the nanosphere. 
In the regime $v \lesssim O(10)$ mm/s, the quantum-mechanical differential cross section Eq.~(\ref{eq:q-diff-scat}) thus reveals the matter-wave diffractions involving interference between different partial waves. Classically, on the other hand, an incident atom is fully characterized by a single angular momentum and there is no interference between atoms with different impact parameters in an incident beam. 

%: elastic and absorption cross sections

Figure~\ref{fig_sigma_Cs} shows elastic and absorption cross sections versus incident atomic velocity. 
Contributions from low-lying $(l \le 8)$ partial waves are also drawn with thin curves. 
The scattering involves various partial waves in the relatively high-energy regime, but 
the contributions from large $l$ gradually decrease as the \Blue{energy} is lowered, and 
eventually only the $s$-wave contribution remains when $v \lesssim 500$~\textmu m/s ($2E/k_B \lesssim O(1)$~nK). 
In the $s$-wave regime, the elastic cross section $\sigma_{\rm el}$ approaches the constant value $4\pi |A_0|^2$. 
This behavior of $\sigma_{\rm el}$ reveals the occurrence of the quantum reflection for the $s$-wave in stark contrast to 
the classical elastic scattering of the vanishing angular momentum. 
As discussed in Sec.~\ref{sec:classical_scat}, the classical elastic cross section shows a fictitious divergence for any velocity associated with the inclusion of infinite impact parameters. If this divergence is properly eliminated, e.g., by using a finite beam, there would be no contribution from the vanishing angular momentum in the classical elastic cross section.

As shown in Fig.~\ref{fig_sigma_Cs} (b), the total absorption cross section $\sigma_{\rm abs}$ and classical absorption cross section $\pi \rho_c^2$ are almost equal, $\sigma_{\rm abs} \simeq \pi \rho_c^2 \propto v^{-4/7}$, for a wide range of velocity $v \gtrsim 500$~\textmu m/s and 
a difference is found only in the $s$-wave regime $v \lesssim 500$~\textmu m/s, 
which is roughly two orders of magnitude smaller than the velocity at which the difference in the elastic differential cross sections starts to emerge. 
This is because the total cross sections are characterized only by diagonal terms of the $S$ matrix and quantum-mechanical interference terms are not involved. The enhanced absorption cross section $\sigma_{\rm abs}$ in the $s$-wave regime is regarded as the manifestation of the quantized impact parameter $l/k$. 

%: rates
%--------------------------------
%: 【figure】elastic and loss rates
\begin{figure}[t]
	\centering
	%\aps{\includegraphics[width=8.7cm,pagebox=cropbox]{fig_10.pdf}}
	\includegraphics[width=8.5cm]{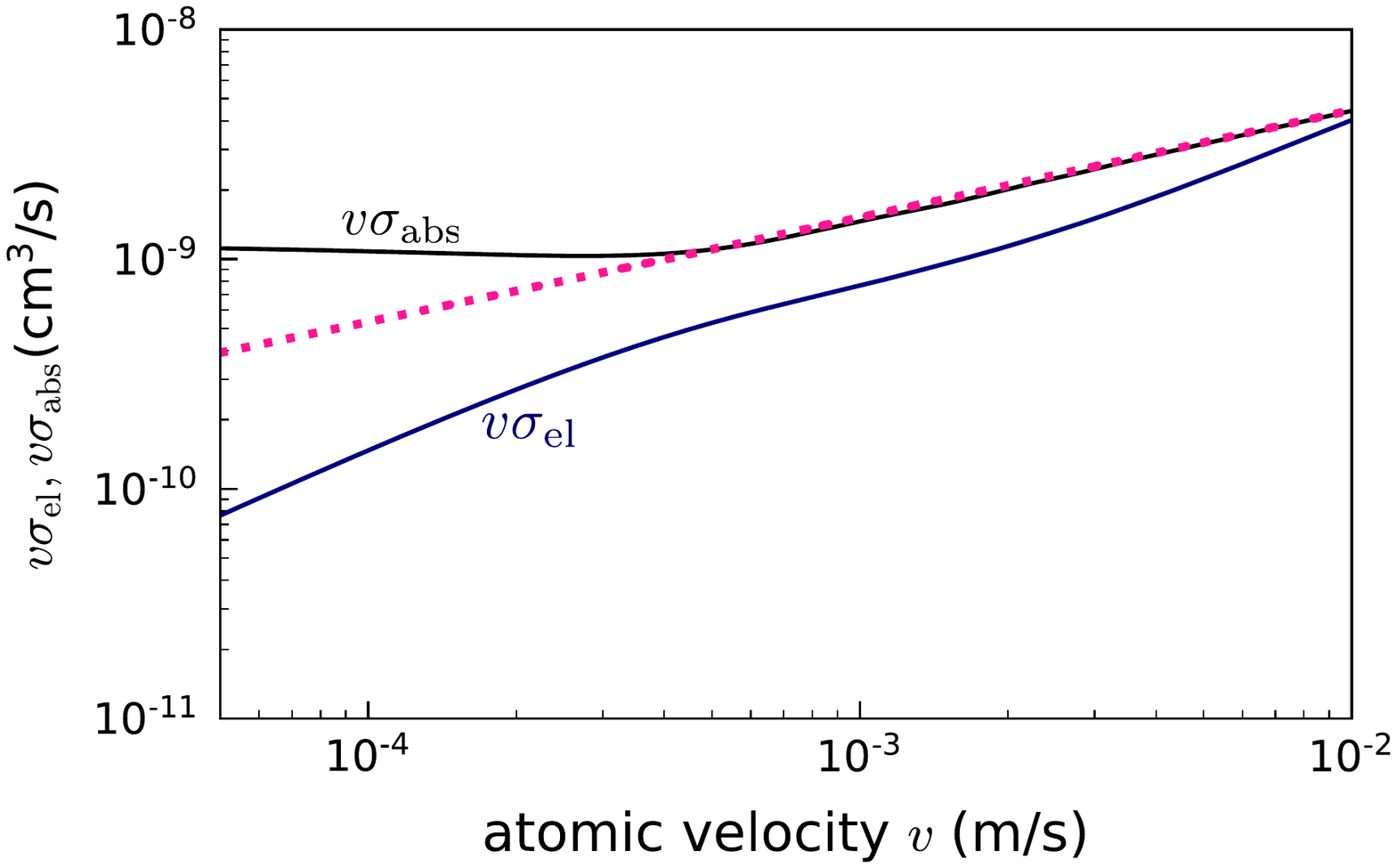}
	\caption{Elastic scattering rate $v\sigma_{\rm el}$ and loss rate $v\sigma_{\rm abs}$ versus incident velocity. Dotted line is the classical loss rate $v \pi \rho_c^2$. In the $s$-wave regime, the quantum-mechanical loss rate is almost constant, while the classical loss rate behaves monotonously.}
	\label{fig:loss-rate}
\end{figure}
%----------------------------------

Another experimentally relevant quantity is the scattering rate, $v \sigma$. We show the elastic scattering rate and the loss rate in Fig.~\ref{fig:loss-rate}. As the \Blue{energy} is lowered, the loss rate associated with the absorption monotonically decreases in $v \gtrsim 500$~\textmu m/s but is nearly constant in the $s$-wave regime, while the classical loss rate continues to monotonically decrease in that regime. 
The elastic scattering rate behaves monotonously for any velocity. 
The optical theorem Eq.~(\ref{optical_theorem_eq}) indicates that the sum of these rates 
$v (\sigma_{\rm el} + \sigma_{\rm abs})= v \sigma_{\rm tot}$ is independent of $v$, which can also be confirmed from Fig.~\ref{fig:loss-rate}.

%: --- high-energy limit 

{\it Relatively high-energy regime} --- 
In a relatively high-energy regime, we find $S_l(k) \approx 0$ for the partial waves of $l \lesssim l_c \approx k \rho_c$. Hence the incoming partial waves of $l \lesssim l_c$ are absorbed onto the surface. 
In this case, the elastic and absorption cross sections have the same value $\sigma_\mathrm{el}^{(l)}(k) = \sigma_\mathrm{abs}^{(l)}(k) = (2l+1) \pi/k^2$, 
as we see from Eqs.~(\ref{eq:sigma_el}) and (\ref{crossection_eq}). By summing partial cross sections up to $l=l_c$, 
the total elastic and absorption cross sections are obtained as 
$\sigma_{\rm el} (k) = \sigma_{\rm abs} (k) = \pi (l_c+1)^2/k^2$. The first term dominantly contributes at high \Blue{energies}, and it almost coincides with the classical absorption cross section $\pi \rho_c^2$.

% ___________________________________________________________________________________________
\subsection{Radius dependence}\label{R-dep}

\begin{figure}[t]
	\centering
	%\aps{\includegraphics[width=8.3 cm,pagebox=cropbox]{fig_11.pdf}}
	\includegraphics[width=8.3 cm]{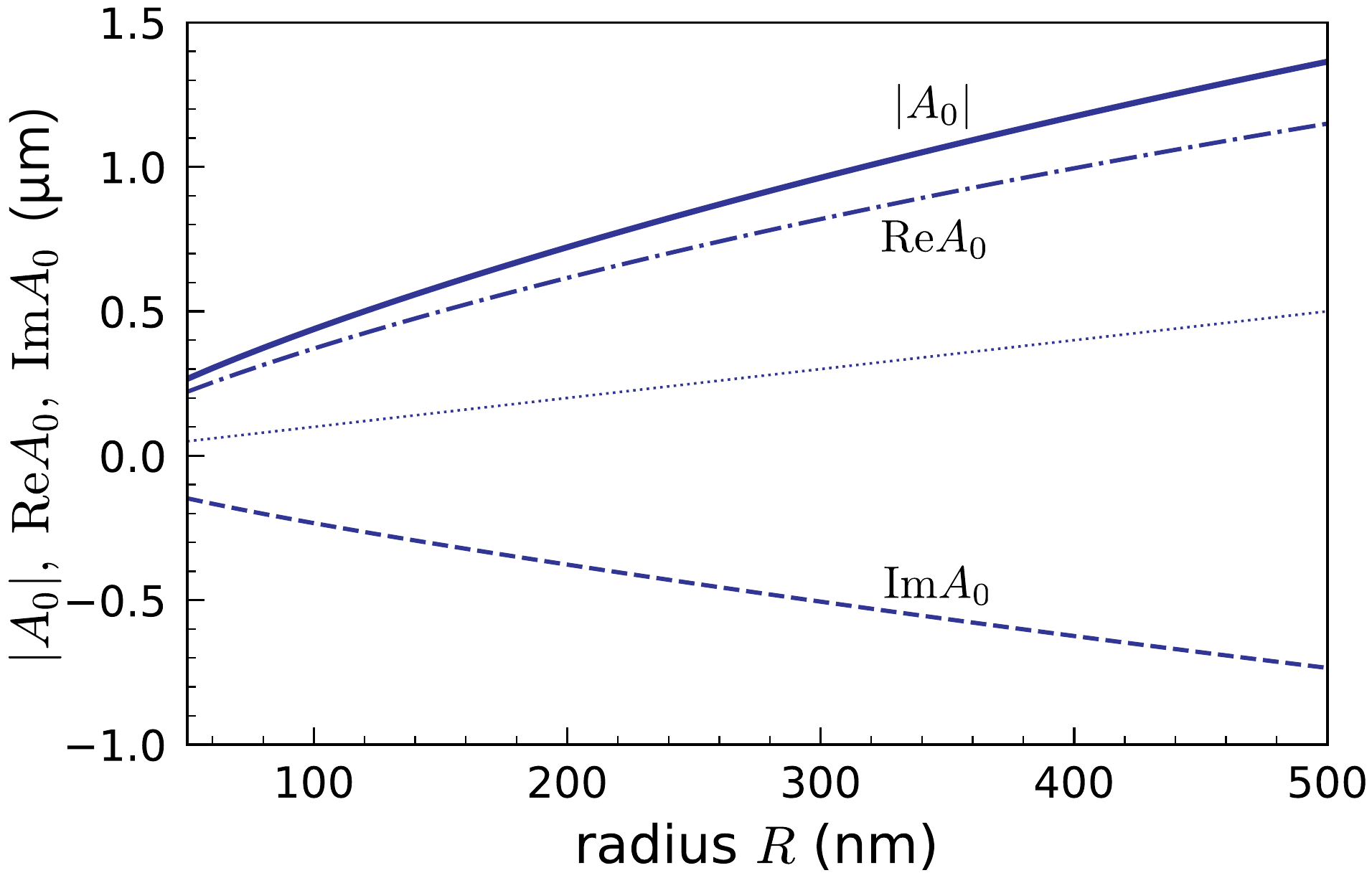}
	\caption{Real, imaginary parts of $s$-wave scattering length $A_0$, and its magnitude $|A_0|$ for a cesium atom. Dotted line shows the geometric radius $R$ of the nanoshpere, i.e., the scattering radius in the case of the hard-sphere potential. }
	\label{fig:A0_vs_R}
\end{figure}

In this subsection we investigate scattering properties as a function of radius $R$ of a nanosphere from 50~nm to 500~nm in the relatively low-energy regime. Figure~\ref{fig:A0_vs_R} shows the complex $s$-wave scattering length $A_0$ of a cesium atom versus $R$. 
If the potential $V(r)$ depends only on $r' = r-R$, the $s$-wave scattering length behaves as $A_0 (R) = A_0 (0) + R$. This is not the case for our potential, since the coefficients $C_6$ and $C_7$ depend on $R$. Nonetheless our scattering radius $|A_0 (R)|$ also monotonically increases versus $R$. 
As compared with the scattering radius of the hard-sphere potential (dotted line in Fig.~\ref{fig:A0_vs_R}), 
that of our dispersion-force potential is found to be more than twice as long as $R$. 

%--------------------------------
%: 【figure】A0
%----------------------------------
%--------------------------------
%: 【figure】cross sections vs R
\begin{figure}[b]
	\centering
	%\aps{\includegraphics[width=8.5cm,pagebox=cropbox]{fig_12.pdf}}
	\includegraphics[width=8.5cm]{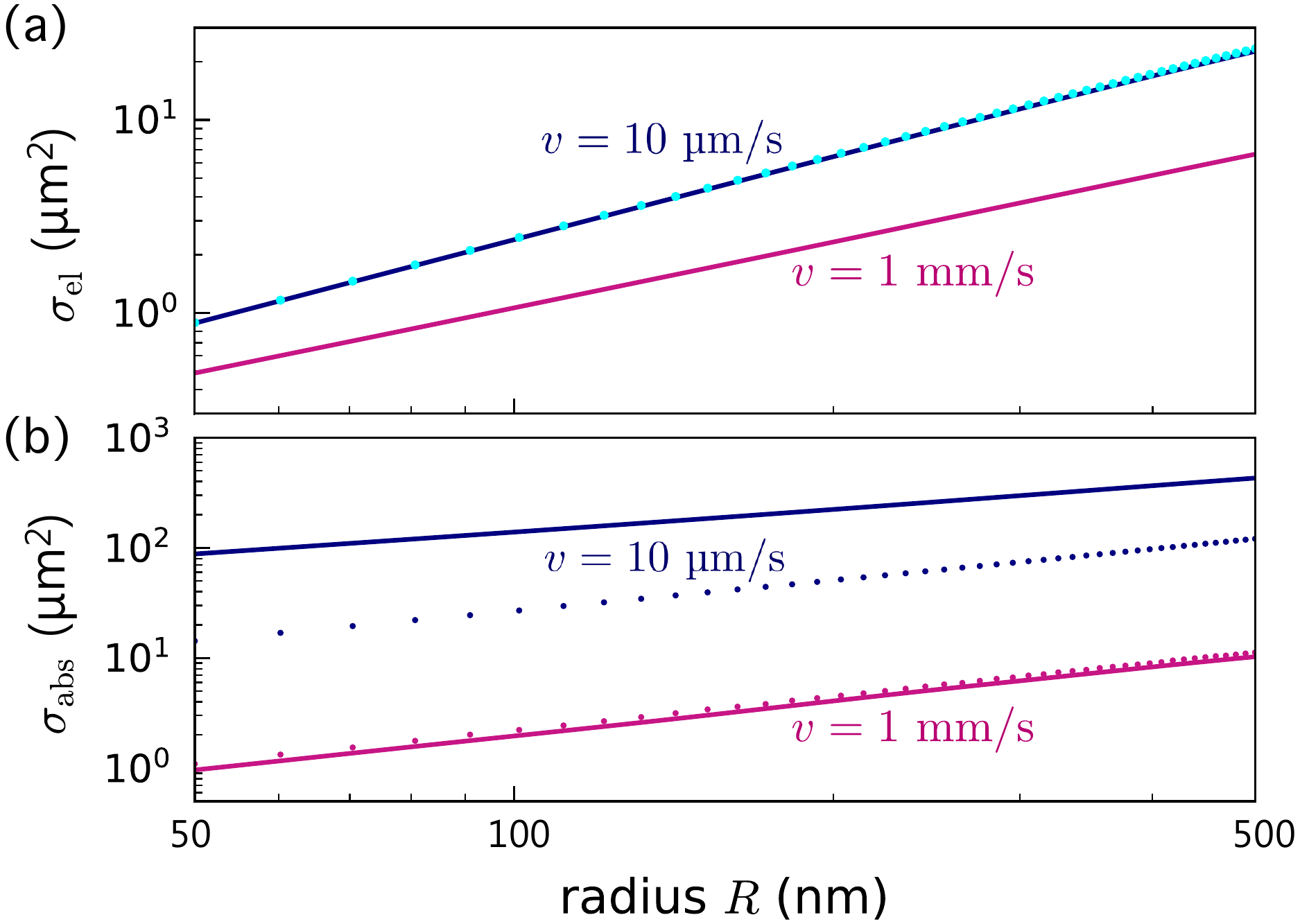}
	\caption{(a) Elastic and (b) absorption cross sections of cesium atoms at $v = 10$~\textmu m/s, and $v = 1$~mm/s. Dotted line in (a) denotes the zero-energy elastic cross section $4\pi |A_0|^2$, and  those in (b) denote 
	the classical absorption cross sections $\pi \rho_c^2$ for each velocity.} 
	\label{fig:sigma_vs_R}
\end{figure}
%----------------------------------

In Fig.~\ref{fig:sigma_vs_R}, we show the elastic cross section $\sigma_{\rm el}$, and the absorption cross section $\sigma_{\rm abs}$ for two incident atomic velocities. 
Both of the elastic and absorption cross sections monotonically increase with respect to $R$, consistent with the intuitive picture of the cross sections. 
When the incident energy is low ($v = 10$~\textmu m/s), the total elastic cross section almost coincides with the zero-energy value $4\pi |A_0|^2$, as shown in Fig.~\ref{fig:sigma_vs_R} (a). 
The absorption cross section $\sigma_{\rm abs}$ is compared with the classical counterpart $\pi \rho_c^2$ in Fig.~\ref{fig:sigma_vs_R} (b). In accordance with the results shown in Fig.~\ref{fig_sigma_Cs} (b), the quantum and classical absorption cross sections agree very well in the relatively high-energy regime, but $\sigma_{\rm abs}$ is larger than $\pi \rho_c^2$ in the low-energy $s$-wave regime. This tendency is seen for the arbitrary radius of the nanosphere. 
Both the quantum and classical absorption cross sections are proportional to $R$ in the velocity regime shown in Fig.~\ref{fig:sigma_vs_R}. This is explained from Eq.~\eqref{eq:rhoc}: The classical cross section consists of three terms including $1/v^{4/7}, R/v^{2/7}$ and $R^2$. Among them, term $R/v^{2/7}$ is dominant, since we consider the nanoscale radius and slow atoms.

% ___________________________________________________________________________________________
\section{Summary}\label{sum}

We studied the low-energy scattering of ultracold cesium and rubidium atoms by a levitated nanosphere of silica glass, 
with special emphasis on the identification of the quantum regime in the potential scattering.

First, we constructed the atom-surface dispersion-force potential from the atom-flat wall potential close to the surface and from the atom-point particle potential at sufficiently long distances. Our potential thus behaves as $V(r) \sim  - (r-R)^{-3}$ in the proximity of the surface, $V(r) \sim - r^{-7}$ at sufficiently long distances, and these smoothly cross over in the middle distances. 
The potential strength was specifically computed for each atom by using the one-oscillator model for the atomic polarizability and the Lorentz model for the dielectric function of the sphere. 

Second, the scattering properties were investigated both classically and quantum mechanically. 
We numerically determined the classical capture range, which was found to be more than one order of magnitude larger than the geometric radius of the nanosphere around the experimentally achievable lowest energy, while it approaches the geometric radius of the nanosphere in the high-energy limit. 
We found that the classical absorption cross section determined from the capture range, and the quantum-mechanical absorption cross section obtained from the $S$ matrix, agree quite well even down to an energy scale of a few nanokelvin in units of temperature.  We also computed loss rates and elastic scattering rates from cross sections, and found good agreement between the quantum and classical loss rates. 
In general, as long as the diagonal elements $S_{l}$ of the $S$ matrix for $l \ge 1$ are concerned, the quantum-mechanical scattering properties are quantitatively similar to the classical ones. 
In other words, in the regime higher than a few nanokelvin, whether the atom is elastically scattered or inelastically lost due to the adsorption, is sorely characterized by the nature of the potential. 
However, in the $s$-wave regime $2E/k_B \lesssim O(1)~$ nK where $S_0 \simeq 1-2\I kA_0$ and $S_{l \ge 1} \approx 1$, the absorption of the $s$ wave is found to be enhanced due to the discreteness of the quantum-mechanical angular momentum. At the same time, the occurrence of the classically-forbidden reflection is identified in the elastic cross section. 
In contrast, we demonstrated that the quantum-mechanical differential cross section of the elastic scattering reveals notable deviations from the classical one in a regime relatively high of the order of a few microkelvin associated with the diffraction, as the manifestation of the wave character of the incident atoms.

The analysis presented in this paper provides insight for the observation of quantum effects in the scattering of ultracold atoms by a dielectric material.

%: Acknowledgments

We thank M.~Bhattacharya and H.~Saito for fruitful discussions. 
This work was partially supported by JST, PRESTO Grant No.~JPMJPR1901, JST, CREST Grant No.~JPMJCR1771, the Matsuo Foundation, and JSPS KAKENHI Grants No. JP16H06017 and No. JP21K03421.  

% ___________________________________________________________________________________________

%\bibliography{bibatomnanosphere}
%\end{document}

% ___________________________________________________________________________________________

\end{document}